\documentclass[lettersize,journal]{IEEEtran}
\usepackage{amsmath,amsfonts}
\usepackage[utf8]{inputenc}
\usepackage{amsmath,amssymb,amsfonts}
\usepackage{array}
\usepackage{algorithm}
\usepackage{algpseudocode}
\usepackage{amsthm}
\usepackage{cite}
\algrenewcommand\algorithmicreturn{\textbf{return}} 
\usepackage{graphicx}
\usepackage[caption=false,font=normalsize,labelfont=sf,textfont=sf]{subfig}
\usepackage{textcomp}
\usepackage{stfloats}
\usepackage{url}
\usepackage{verbatim}
\usepackage{graphicx}
\usepackage{cite}
\usepackage{booktabs}
\begin{document}

\title{A Timing-Anomaly Free Dynamic Scheduling on Heterogeneous Systems}

\author{ Yixuan Zhu, Yinkang Gao, Lei Gong~\IEEEmembership{Member,~IEEE,} Binze Jiang, Xiaohang Gong, Zihan Wang, Cheng Tang,  Wenqi Lou, Teng Wang, Chao Wang~\IEEEmembership{Senior Member,~IEEE,} Xi Li, Xuehai Zhou
        % <-this % stops a space
\thanks{Yixuan Zhu, Yinkang Gao, Binze Jiang, Xiaohang Gong, Zihan Wang, Cheng Tang, Lei Gong, Wenqi Lou, Teng Wang, Chao Wang, Xi Li, and Xuehai Zhou are with the University of Science and Technology of China (USTC), Hefei, China, and also with Suzhou Institute for Advanced Research, USTC, Suzhou, China 
(e-mail: \{zhuyixuan, gaoyinkang, sazb, gxh2018, jiangbinze, wangzh196, tangcheng\}@mail.ustc.edu.cn; \{leigong0203, louwenqi, wangt635, cswang, llxx, xhzhou\}@ustc.edu.cn).}
}

\markboth{Transactions on Computer-Aided Design of Integrated Circuits and Systems}%
{Shell \MakeLowercase{\textit{et al.}}: A Sample Article Using IEEEtran.cls for IEEE Journals}


\maketitle

\begin{abstract}
Heterogeneous systems commonly adopt dynamic scheduling algorithms to improve resource utilization and enhance scheduling flexibility. However, such flexibility may introduce timing anomalies, wherein locally reduced execution times can lead to an increase in the overall system execution time. This phenomenon significantly complicates the analysis of Worst-Case Response Time (WCRT), rendering conventional analysis either overly pessimistic or unsafe, and often necessitating exhaustive state-space exploration to ensure correctness.

To address this challenge, this paper presents the first timing-anomaly-free dynamic scheduling algorithm for heterogeneous systems, referred to as Deterministic Dynamic Execution. It achieves a safe and tight WCRT estimate through a single offline simulation execution. The core idea is to apply deterministic execution constraints, which partially restrict the resource allocation and execution order of tasks at runtime. Based on a formally defined execution progress model for heterogeneous system scheduling, we prove the correctness of the proposed execution constraints and their ability to eliminate timing anomalies. Furthermore, we propose two methods to generate execution constraints. The first method derives execution constraints directly from the execution traces produced by existing scheduling algorithms. The second method is a heuristic-based approach that constructs execution constraints, enabling further reduction of the WCRT. Experimental results on synthetically generated DAG task sets under various system configurations demonstrate that, compared to traditional dynamic scheduling algorithms, our approach not only eliminates timing anomalies but also effectively reduces both the WCRT and response time jitter.
\end{abstract}


\section{Introduction}
Heterogeneous multi-core~\cite{Heterogeneous_multicore_1, Heterogeneous_multicore_2} architectures have gained widespread attention in emerging embedded systems with high computational demands and real-time requirements, due to their high performance and flexible resource composition. Compared to homogeneous systems, heterogeneous platforms integrate diverse types of processing units, enabling higher degrees of parallelism and improved resource utilization. However, the heterogeneity of resources and variations in execution behavior introduce significant challenges for task scheduling and Worst-Case Response Time (WCRT) analysis~\cite{WCRT_alysi}.

% In heterogeneous multicore systems, task model is typically represented as a typed DAG~\cite{typedDAG, Han1, Han2}. 
In heterogeneous multicore systems, tasks are always scheduled by either static or dynamic scheduling. 
Static scheduling~\cite{staticSch,staticSchHETF,staticSchHETFnew} determines the fixed mapping between tasks and specific instances of processing units at design time. 
It enables precise WCRT analysis but lacks flexibility, often resulting in poor resource utilization and parallelism. Dynamic scheduling~\cite{dySCh1, dySCh2} makes decisions at runtime, making it better suited to the resource heterogeneity and execution variability of heterogeneous platforms. 

However, dynamic scheduling may introduce \textit{timing anomaly (TA)}~\cite{818824, graham1969bounds, reineke2006definition}. It is a situation where a locally shorter execution time of tasks leads to a globally longer response time of the system, which may lead to two problems.
First, it makes the WCRT analysis unsafe, which is estimated via offline simulation using worst-case execution time (WCET)~\cite{WCET1} may be exceeded by the result of actual online execution. 
Second, it complicates accurate WCRT analysis. It requires exhaustive exploration of all execution time combinations across heterogeneous units, leading to a state space explosion. 
Therefore, designing a TA-free dynamic scheduling algorithm, which can produce an accurate and safe WCRT, is critical for heterogeneous real-time platforms.

Existing research on TAs mainly focuses on homogeneous systems. 
Works~\cite{TAFree1, TAFree2, TAFree3, TAFree4} primarily propose a class of dynamic scheduling algorithms that are formally proven to be TA-free. Specifically, they eliminate TAs at the scheduling level by constraining either the degree of system parallelism or the online execution order of tasks. For such algorithms, a single offline simulated execution based on all tasks' WCETs (all-WCETs) can yield a safe and accurate WCRT. 
However, extending these methods to heterogeneous systems is non-trivial due to several challenges: First, dynamic scheduling on heterogeneous systems introduces greater uncertainty. Since a task can be assigned to different heterogeneous processing units, it may have multiple execution times. 
As a result, these methods used in homogeneous systems may not be sufficient to eliminate TAs in this scenario. Therefore, it is necessary to develop new methods and design a corresponding TA-free dynamic scheduling algorithm. Second, it is challenging to prove that they can guarantee to eliminate TAs. This is because it requires verification that the method remains TA-free across all systems, including different task sets and heterogeneous resource configurations, which cannot be validated purely through experimental simulations. Therefore, a rigorous formal proof is the only viable approach to establishing this guarantee.

To address the issue mentioned above, this paper proposes a formally verified execution constraint capable of eliminating TAs for non-preemptive heterogeneous systems. Building on this, we introduce the first TA-free dynamic scheduling algorithm—Deterministic Dynamic Execution (DDE).
Specifically, we begin by constructing an execution-progress-based model to 
describe the scheduling in heterogeneous systems. Based on it, we define a strict form of TA that encompasses the corresponding normal TAs. Therefore, if a system is free of strict TAs, it is also free of normal TAs.
We then design the deterministic execution constraint, which can eliminate the strict TAs. This constraint enforces partial restrictions on resource allocation and execution order. 
We formally prove that this constraint ensures the monotonicity of execution progress, which further guarantees the system's freedom from strict TAs.
Next, we design a TA-free dynamic scheduling algorithm by this constraint.
Then, we propose two methods for obtaining execution constraints. The first one is a trace-based approach that extracts constraints from the traces of simulated execution. The second method is a heuristic algorithm for constraint generation, which can further reduce the WCRT.
Experimental results demonstrate that our method not only eliminates TAs but also reduces the system's WCRT by an average of 5\%-25\%, with a best-case reduction of 72\%, and response time jitter by 7\%-9\% across various test scenarios. 
% Although the average response time increased by 4.2\% in random systems, in systems with TA, the system's average response time still decreased by 1.2\%.
% Although the average response time increased by 4.2\% in random systems with or without TAs, it decreased by 1.2\% in systems with TAs.
Although the average response time increased by 4.2\% when considering all systems (including those with and without TAs), it decreased by 1.2\% when focusing specifically on systems that exhibit TAs. The main contributions of this paper are as follows:
\begin{itemize}
    \item Establishing an execution progress model for the heterogeneous systems to capture timing behaviors during dynamic scheduling.   
    \item Defining a strict TA by execution progress model with mulit local reductions, which can encompass normal TAs.
    \item Proposing the first (normal) TA-Free dynamic scheduling algorithm by our 
    execution constraint with a formal proof of strict TA-free for heterogeneous systems. 
    \item Giving two methods for obtaining execution constraints, which can significantly reduce the WCRT.
    \item Achieving Joint elimination of TA and multiple metrics optimization, including WCRT and response time jitter.
\end{itemize}

\section{PRELIMINARY \& MOTIVATION}
We first introduce the heterogeneous system, task model, and WCRT analysis. 
Then, we present the motivation for this work by a case of timing anomaly in a heterogeneous system.

\subsection{Heterogeneous System}
\label{ChapterII_Sec_A}

In this work, we consider a heterogeneous computing platform consisting of multiple types of processing units, such as CPUs, GPUs, and domain-specific accelerators~\cite{ACC1, NPU1, houssam2020hpc}. Due to their inherently diverse performance characteristics, tasks are constrained to execute only on compatible subsets of resources, and their execution times may vary across eligible units.
For example, a Xavier-based platform such as NVIDIA Pegasus integrates 8 CPU cores, 4 GPUs (2 discrete GPUs, 2 integrated GPUs), 2 Deep Learning Accelerators (DLAs), and 2 Programmable Vision Accelerators (PVAs). On this platform, a deep learning inference task can run on CPUs, GPUs, or DLAs, each exhibiting distinct execution-time characteristics~\cite{houssam2020hpc}.

To uniformly denote the processing units, we represent the instance of a processing unit as $\mathcal{A}_{\mathcal{T}}^{\mathcal{I}}$. $\mathcal{A}$ specifies the \emph{architecture} (e.g., CPU), and $\mathcal{T}$ is a \emph{type} within a specific architecture, capturing differences in processing capability (e.g., $\text{CPU}_1$ for a high-performance core and $\text{CPU}_2$ for a low-power core). 
The superscript $\mathcal{I}$ denotes the \emph{instance index} of the architecture $\mathcal{A}$ and type $\mathcal{T}$, so that $\text{CPU}_1^0$ and $\text{CPU}_1^1$ refer to two identical instances with different indices of $\text{CPU}_1$.

\subsection{Task Model}
\label{lab:taskmodel}
This paper adopts a \textit{multi-typed DAG} to model heterogeneous systems. A multi-typed DAG task set $\mathbb{T}$ is represented as $G = (V, E, M, \Gamma)$, where: 
$V$ is the set of vertices, each representing a task;
$E \subseteq V \times V$ is the set of directed edges, where $(u, v) \in E$ indicates that task $u$ must finsh before task $v$ can begin (i.e., $u$'s finish triggers $v$'s start);
$M: V \rightarrow 2^{CU} \setminus \emptyset$ is a mapping from a task $v$ to the set of its corresponding \textit{processing unit types}, i.e., $M(v) \subseteq {PUT}$, where $PUT$ is the set of all processing unit types \textit{$put$} in the system;
$\Gamma: V \times {PUT} \rightarrow \mathbb{I}_{>0}$ is the \textit{execution time interval function}, where $\Gamma(v, put) = [BCET(v, put), WCET(v, put)]$ denotes the execution time interval of task $v$ on a processing unit model $put \in M(v)$. Modeling execution time as an interval reflects internal workload variations within the task, leading to runtime on a given processing unit instance ranging from its Best-Case Execution Time (BCET) to its WCET.

\begin{figure}
    \centering
    \includegraphics[width=0.85\linewidth]{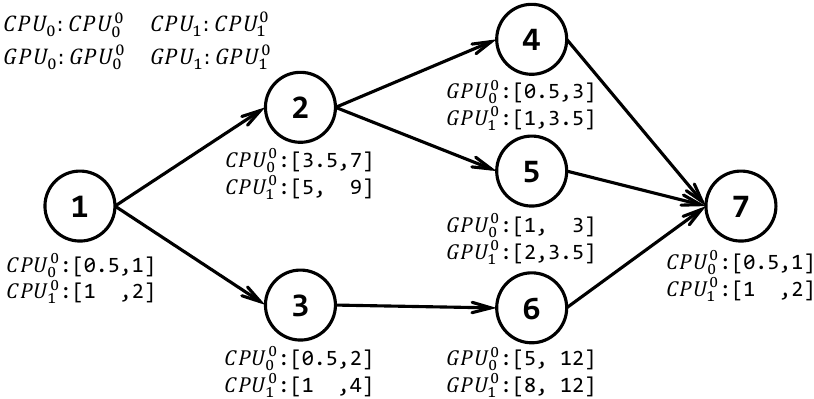}
    % \vspace{-0.2cm}
    \caption{An example of a multi-typed DAG with seven nodes (tasks) and corresponding processing units with execution times for each node.}
    \label{fig:DAG1}
    % \vspace{-0.6cm} 
\end{figure}

As shown in Fig.~\ref{fig:DAG1}, it depicts an example of a multi-typed DAG. The upper-left corner shows all available processing-unit instances, including two CPU types ($\text{CPU}_0, \text{CPU}_1$) and two GPU types ($\text{GPU}_0, \text{GPU}_1$), each with an instance. For each node, the set of eligible processing units and the corresponding execution times are specified below the node.

Nodes in the DAG with no incoming edges are referred to as \textit{source nodes}, denoted by $v_{\text{src}}$, and nodes with no outgoing edges are referred to as \textit{sink nodes}, denoted by $v_{\text{sink}}$. If a DAG contains multiple source or sink nodes, a virtual source node is added to connect all source nodes, and a virtual sink node is added to connect all sink nodes. These virtual nodes have zero execution time and do not occupy computing resources. The dependency relationships between tasks are defined by directed edges. A task can enter the \textit{\textbf{ready}} state only after all its predecessor tasks have completed. 

For a task $v_i$ of a node, let its start time be denoted as $s_i$ and its finish time as $f_i$. The response time of task $v_i$, denoted as $RT(i)$, is defined as the time interval from the start time of the source node to the finish time of $v_i$, i.e., $f_i - s_{\text{src}}$. The response time of the DAG system $\mathbb{T}$ is given by $RT(v_{sink})$.

\subsection{WCRT Analysis Based on Simulated Execution }
Under a scheduling algorithm, the WCRT of a DAG is the maximum response time over all possible scheduling results of all tasks in the system. Since the completion of the DAG is determined by its sink nodes, the  WCRT can be simplified to the maximum response time of the sink node. Formally, let $\mathbb{S}$ denote the set of all feasible schedules generated by a given scheduling algorithm for a DAG and $RT_\mathcal{S}$ denotes the response time under a schedule $\mathcal{S}$.
The WCRT is defined by Eqn.~\ref{WCRTbySim}. It is typically obtained by all-WCET simulated execution offline, in which each task is assumed to execute with its WCET. 
\begin{equation}
\mathrm{WCRT}
=
\max_{\mathcal{S} \in \mathbb{S}}
\; RT_\mathcal{S}(v_{\text{sink}})
% \tag{0}  
\label{WCRTbySim}
\end{equation}

\begin{figure}
    \centering
    \includegraphics[width=\linewidth]{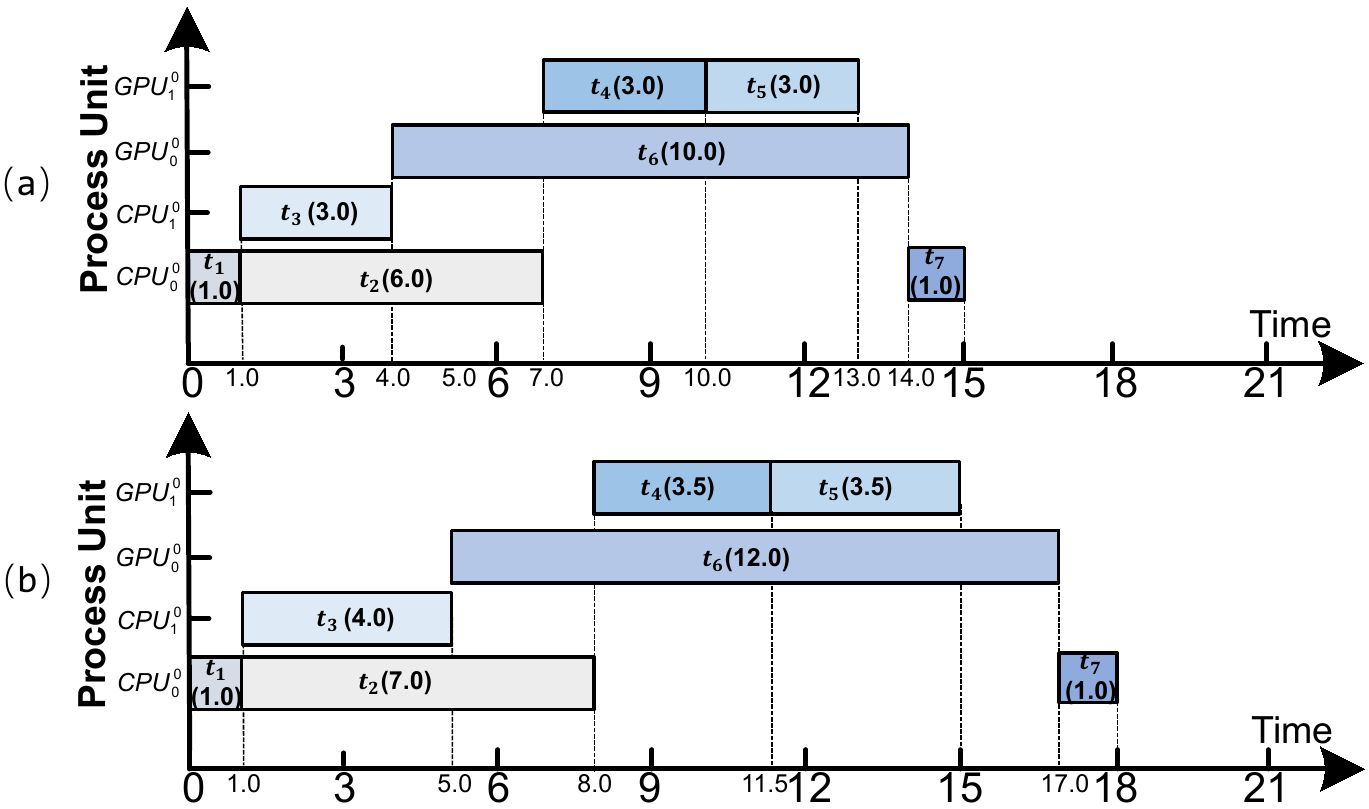}
    \caption{Two schedules of the DAG in Fig. 1 under HBFS: (a). an arbitrary online execution with response time 15, and (b). all-WCETs offline simulated execution with response time 18, which is always considered as WCRT. }
    \label{fig:twoExectionCase}
\end{figure}

For example, consider the system in Fig.~\ref{fig:DAG1}, where we apply the HBFS (Heterogeneous Breadth-First Scheduling) algorithm. It is a heterogeneity-aware extension of the homogeneous BFS~\cite{TAFree1}, which prioritizes tasks closer to the source node, with tasks having smaller identifiers given higher priority when distances are equal. It prefers allocating tasks to faster processing units. Fig.~\ref{fig:twoExectionCase}(b) shows the offline simulation results under all-WCETs assumptions with WCRT 18.

However, during online execution, task execution times are inherently uncertain due to variations in input workloads, typically fluctuating between their BCET and WCET on a given processing unit. This variability of execution time induces uncertainty in the system’s runtime state, including the contents of ready queues and the availability of computing resources. Consequently, the set of processing units available to a task at runtime becomes unpredictable, which in turn further amplifies variability of execution time and ultimately affects the system’s response time.
As shown in Fig.~\ref{fig:twoExectionCase}(a), which presents an online execution instance of the DAG in Fig.~\ref{fig:DAG1} under the HBFS policy, some tasks execute with durations shorter than their WCETs. The system’s observed response time in this schedule is 15, which is shorter than the WCRT of 18 obtained from all-WCETs execution in Fig.~\ref{fig:twoExectionCase}(b).

\subsection{Motivation}
\label{mov}
% \section{Motivation}
However, the above method does not always yield a safe WCRT. Its validity relies on the assumption that using local maximums can lead to the global maximum. However, for non-preemptive dynamic schedules like HBFS, which adopt a locally greedy approach, this assumption often does not hold. A counterintuitive phenomenon may occur: a task executing for less than its WCET during online execution may actually lead to a bigger system response time than obtained when all-WCETs execution. This phenomenon is known as TA.

\begin{figure}
    \centering
    \includegraphics[width=1\linewidth]{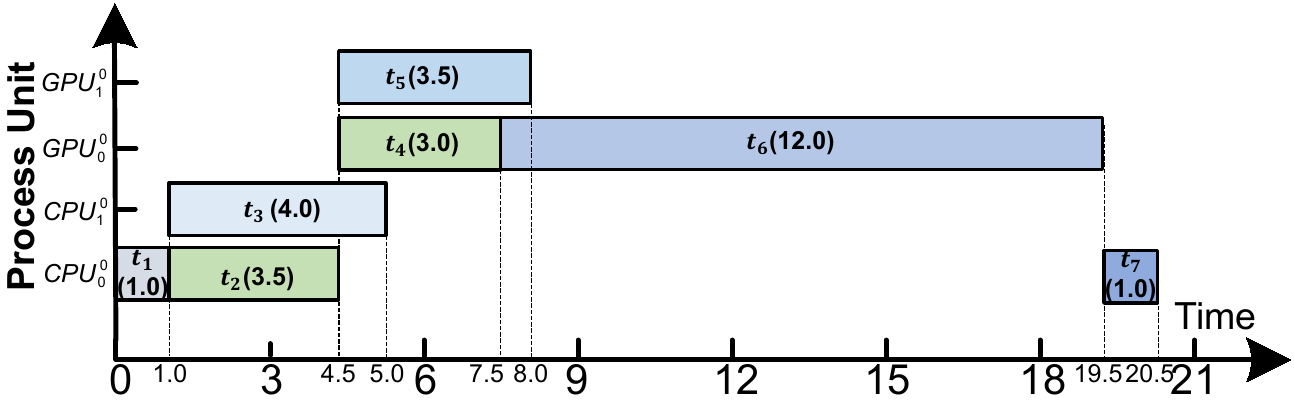}
    \caption{An example of a TA for the DAG shown in Fig.~\ref{fig:DAG1} under the HBFS scheduling, where the execution time of the tasks highlighted in green is shorter than the WCET in Fig.~\ref{fig:twoExectionCase}(b), resulting in a response time of 20.5. It exceeds the WCRT of 18 obtained from the all-WCETs execution in Fig.~\ref{fig:twoExectionCase}(b).}
    \label{fig:TA-case}
\end{figure}

Here, a TA is illustrated by the example in Fig.~\ref{fig:TA-case}. It can be observed that the response time is $20.5$, which is bigger than $18$ obtained by offline simulated execution by all-WCETs. 

\begin{figure}
    \centering
    \includegraphics[width=1\linewidth]{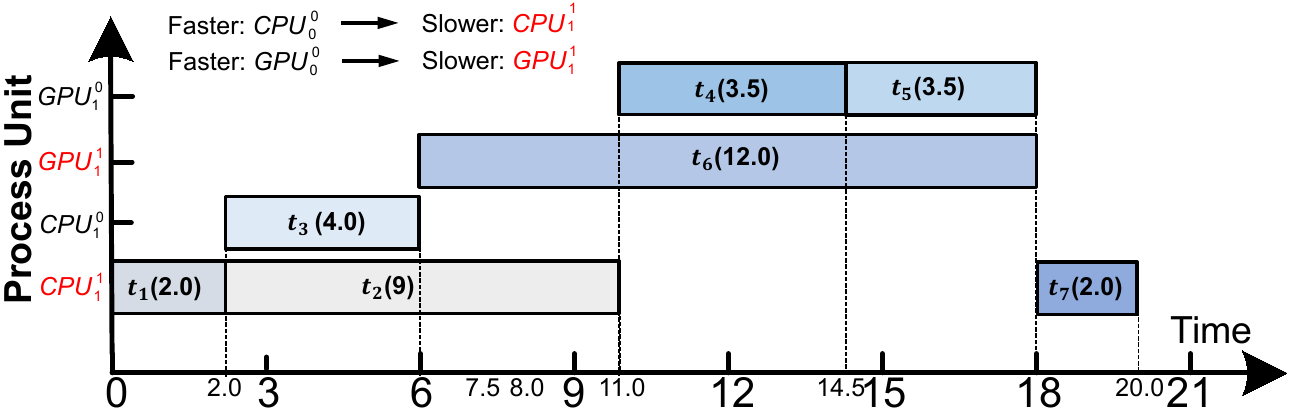}
    \caption{A conservative simulated execution by transforming the faster processing units in Fig.~\ref{fig:DAG1} to the slowest type within the same architecture and performing an all-WCETs estimation, the resulting WCRT may still not be safe. The obtained WCRT is 20 remains smaller than the 20.5 in Fig.~\ref{fig:TA-case}.}
    \label{fig:TA-conservative}
\end{figure}

% Even with more conservative offline execution to estimate WCRT, TA may still occur in the system. In heterogeneous systems, during offline simulation, a task might be assigned to a high-performance processing unit, but this does not represent the worst-case scenario for that task in the actual system. To enhance the safety of the analysis, we adopt a conservative approach in the offline phase by modeling each type of processing unit as the slowest unit within that type, ensuring that tasks are always executed under worst-case conditions. Based on this conservative assumption, the computed $RTA$ is 20, which is still lower than the 20.5 observed in Fig~\ref{fig:TA-case}, indicating that a TA still occurs.
Even with conservative offline simulation, the WCRT obtained may still be unsafe, and TAs may occur. In heterogeneous systems, as shown in Fig.~\ref{fig:twoExectionCase}(b), a task might be assigned to a type of faster processing unit, but this does not represent the worst-case scenario for that task in the actual system. For example, tasks 1, 2, and 7 in Fig.~\ref{fig:twoExectionCase}(b) are assigned to the faster type $CPU_0$, but in reality, they may run on a slower type $CPU_1$ with a bigger WCET, which is not reflected in this simulation. 
% To enhance the safety of the analysis, we model all processing
% units within the same architecture at reduced speed, representing the slowest type, and then perform an all-WCET simulation, ensuring that tasks are always executed under worst-case conditions. 
To enhance the safety of the analysis, we model all processing units within the same architecture as the slowest type in that architecture. We then perform an all-WCETs simulation to ensure that tasks are always executed under worst-case scenario.
As shown in Fig.~\ref{fig:TA-conservative}, the conservatively estimated response time is 20, still lower than the 20.5 from Fig.~\ref{fig:TA-case}, indicating that a TA still occurs.

When a TA occurs, tasks execute for less than their WCETs, leading to a longer global response time. This indicates that the \textit{reduction} of response time from reduced tasks' execution times, due to lighter workloads or the selection of faster processing units, is insufficient to offset the \textit{increase} of response time caused by reduced system parallelism. However, whether a TA occurs depends on the relative magnitudes of these gains and losses at the end of the system's execution. The \textit{reduction} and \textit{increase} values can dynamically change during execution, and their relationship is difficult to quantify, especially under complex parallelism and dynamic heterogeneous processor selection. Consequently, it is not easy to precisely eliminate TAs in heterogeneous systems, including establishing sufficient and necessary conditions and certifications, which are the challenges mentioned before.

% When a TA occurs, tasks execute for less than their WCETs, leading to a longer global response time. This indicates that the performance gain (\textit{increase}) from reduced tasks' execution times, due to lighter workloads or the selection of faster processing units, is insufficient to offset the performance loss (\textit{reduction}) caused by reduced system parallelism. However, whether a TA occurs depends on the relative magnitudes of these gains and losses at the end of the system's execution. The \textit{reduction} and \textit{increase} values can dynamically change during execution, and their relationship is difficult to quantify, especially under complex parallelism and dynamic heterogeneous processor selection. Consequently, it is not easy to precisely eliminating TAs in heterogeneous systems, including establishing sufficient and necessary conditions and certifications, which are the challenges mentioned before.

To address this, we are prepared to adopt a conservative approach to eliminate TAs. Specifically, we ensure that the \textit{reduction} from reduced task execution times is always greater than the \textit{increase} due to reduced system parallelism, maintaining a negative difference throughout the entire execution of the system. This guarantees that the end state of the system is free of TAs. It is important to note that this method provides a sufficient, though not necessary, means of eliminating TAs. And it is based on deterministic execution constraints, whose TA-free property will be proved later via monotonicity, using our execution-progress model and a strict definition of TAs.
% be proven by our execution progress model and later through monotonicity.

\section{System Modeling}
% We construct the task execution progress model and then the system execution progress model.
We construct the task and system execution progress model based on a finite state machine. Building on this, we give the definition and method to compare the relationship of progress.

\subsection{Task Execution Progress Model}
The task execution progress model consists of the task's current execution stage and its remaining execution time in this stage. The model is designed to characterize the relative magnitude of the remaining execution time required for the same task across two different execution instances (schedules).

The execution stage of a task is represented by \( s \), where \( s \in S \) and $S$ is defined by the task's execution flow as follows.
\begin{equation}
 S = \{ \texttt{Block}, \texttt{Ready}, \texttt{Exec}, \texttt{Finish} \}
% \tag{1}  
\label{Eq_stage}
\end{equation}

Specifically, the \textbf{Block} stage indicates that the task has unfinished predecessors and cannot yet be scheduled. The \textbf{Ready} stage means that all dependencies have been resolved, and the task is eligible for scheduling. 
% The \textbf{Exec} stage indicates that the task has been dispatched to a computing resource and is currently executing; it also implicitly captures the specific processing unit from the Sec.~\ref{ChapterII_Sec_A} assigned to the task. 
The \textbf{Exec} stage indicates that the task has been dispatched to a processing unit and is currently executing. It also implicitly specifies the instance of the assigned unit. For example, when a task's Exec is $\text{CPU}_1^0$ as from Sec.~\ref{ChapterII_Sec_A}, it means the task is in the execution stage and has been assigned to processor instance $\text{CPU}_1^0$.
The \textbf{Finish} stage indicates that the task has completed execution. 

Let \( t \in \mathbb{N}_0 \) denote the remaining execution time of a task in its current stage. 

The execution progress of a task is denoted by a \( p \in P \), where \( P:= S \times \mathbb{N}_0 \) represents all possible progress states. We define a partial order \( \sqsubseteq_P \) over execution progress such that \( p_1 = (s_1, t_1) \sqsubseteq_P p_2 = (s_2, t_2) \) indicates that execution progress of \( p_1 \) is no greater (less or equal) than  \( p_2 \), either because the task is in an earlier stage or, when in the same stage, has a smaller remaining execution time (Eqn.~\ref{Eq_taskProgressOrder}). The relation \( \sqsubseteq_S \) denotes a partial order over stages, depending on the system with $
\texttt{Block} \sqsubset_S \texttt{Ready} \sqsubset_S \texttt{Exec} \sqsubset_S \texttt{Finish} $.

\begin{equation}
p_1 \sqsubseteq_P p_2  \Leftrightarrow s_1 \sqsubseteq_S s_2 \lor (s_1 = s_2 \land t_1 \geq t_2)
% \tag{1}  
\label{Eq_taskProgressOrder}
\end{equation}

\subsection{System Execution Progress Model}
\begin{figure}
    \centering
    \includegraphics[width=1\linewidth]{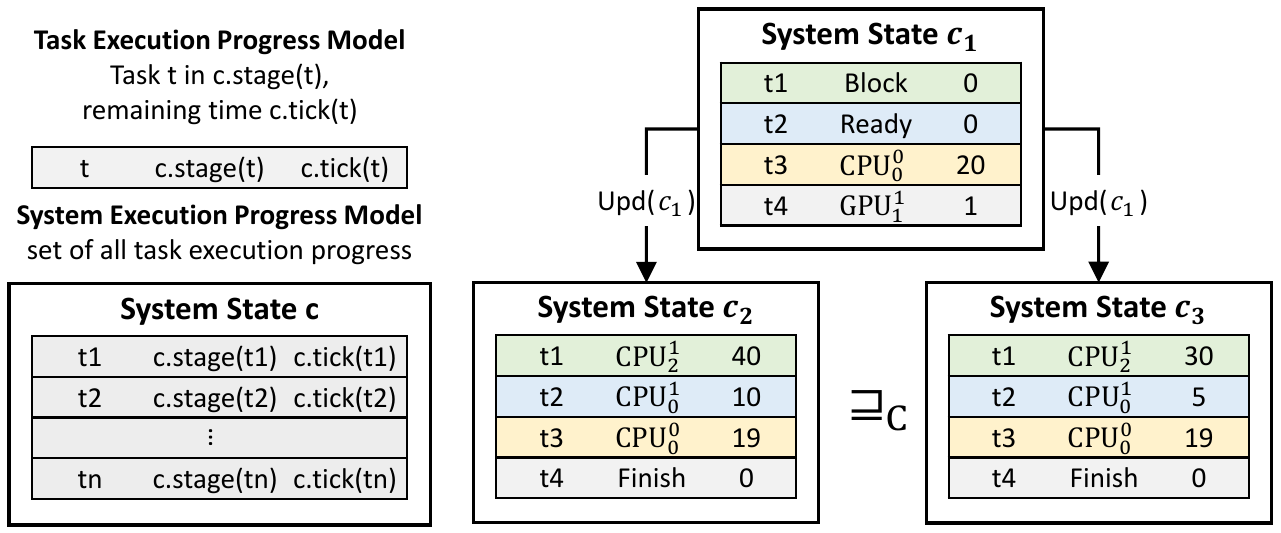}
    \caption{System execution progress model and partial order of progresses }
    \label{fig:implementation-progres}
\end{figure}

We construct the system execution progress model and its state transitions based on the task execution progress model. The \( \mathbb{T} \) denotes the set of all tasks in the system. The system state can be represented as a mapping from each task to its corresponding execution progress, formally defined as \( C:= \mathbb{T} \rightarrow P \). A system state is denoted by \( c \in C \), where for any task \( t \in \mathbb{T} \), \( c(t) \in P \) indicates the execution progress of task \( t \) under that state $c$. As illustrated in Fig.~\ref{fig:implementation-progres}, the system state is characterized by the execution progress of its four tasks.

We define a partial order \( c \sqsubseteq_C c' \) between two system states $c$ and $c'$, meaning that the execution progress of \( c \) is no greater than \( c' \). This holds if and only if, for every task \( t \in \mathbb{T} \), the execution progress in \( c \) satisfies \( c(t) \sqsubseteq_P c'(t) \), as formalized in Eqn.~\ref{Eq_systemProgressOrder}. For example, the system state \( c_2 \) is no greater than \( c_3 \) in terms of execution progress in Fig.~\ref{fig:implementation-progres}, 
\begin{equation}
c\sqsubseteq_C c' \Leftrightarrow \forall t\in \mathbb{T}, c(t) \sqsubseteq_P c'(t)
% \tag{2}
\label{Eq_systemProgressOrder}
\end{equation}

\subsection{System Execution State Transition}
\label{systemEST}
The system execution can be modeled as a state transition function \( upd: C \rightarrow C \). Each application of \( upd \) corresponds to one minimal scheduling cycle, during which the global system state is updated based on the task statuses and resource availability. As illustrated in Fig.~\ref{fig:implementation-progres}, this process depicts a transition from state \( c_1 \) to another state. 

% In heterogeneous systems, the minimal time unit is defined as a discrete scheduling step, representing a unified time granularity across different computing resources, ensuring that all tasks progress by discrete integer steps. 

$upd$ is the transition function as Eqn.~\ref{Eq_upd:udp}. It is constructed from $tran(t)$, $stage'(t)$, $tick'(t)$, and $allocTime(t)$, as represented by Eqn.~\ref{Eq_tran}, Eqn.~\ref{Eq_stage_}, Eqn.~\ref{Eq_tick_}, and Eqn.~\ref{Eq_altime}, respectively. And the $\lambda t \in \mathbb{T}$ denotes an arbitrary task $t$ in $\mathbb{T}$.
\begin{equation}
upd:= \lambda t\in \mathbb{T} \left\{
                \begin{array}{ll}
                    (stage'(t), allocTime(t)) : tran(t)\\
                    (stage(t), tick'(t)) : \text{otherwise} \\
                \end{array}
            \right.
% \tag{3}
\label{Eq_upd:udp}
\end{equation}

$stage(t)$ is current stage of $t$ and an element from Eqn.~\ref{Eq_stage}.

$tran(t)$  determines whether task $t$ is ready to transition to the next stage, returning true if the corresponding conditions are met, as defined in Eqn.~\ref{Eq_tran}.

$tick(t)$ is the remaining time for the currtent stage $stage(t)$.
\begin{equation}
\begin{aligned}
tran(t) := & (tick(t) \leq 1) \\
& \land (stage(t) = \texttt{Block} \Rightarrow depComp(t)) \\
& \land (stage(t) = \texttt{Ready} \Rightarrow resAble(t)) \\
& \land (stage(t) \in \texttt{Exec} \Rightarrow \text{true}) \\
& \land (stage(t) = \texttt{Finish} \Rightarrow \text{true})
\end{aligned}
\label{Eq_tran}
\end{equation}

$depComp(t)$ returns true if all predecessor tasks of 
$t$ has finished, e.g, the dependencies in the DAG have been resolved.

% $resAble(t)$ returns true if a suitable processing unit is available for $t$ at the next state, which is determined by:
% \begin{equation}
% \begin{aligned}
%                 resAble(t) := &resNum(t) > (coutHiCompRes(t) \\ 
%                             & +  nextStageStillUse(stage'(t))
%             \end{aligned}
% \label{Eq_resAble}    
% \end{equation}

$resAble(t)$ returns true if there exists an available processing unit for $t$ at the next state, determined by Eqn.~\ref{Eq_resAble}. And $M(t)$ is the type of suitable processing unit defined in Sec.~\ref{lab:taskmodel}.
\begin{equation}
\begin{aligned}
& resAble(t) :=
\exists\, res \in M(t) :
resNum(res) 
> \\ 
& \ hiTaskNum4Res(res, t) + nextStageStillUse(res)
\end{aligned}
\label{Eq_resAble}
\end{equation}

% $resNum(t)$ denotes the number of uitable processing units for the task $t$ in the system, composed of the total number of instances of the uitable types for $t$.

$resNum(res)$ is the number of instances for the type $res$.

% $resNum(res)$ is the number of uitable instances of processing units with types from $M(t)$, and $M(t)$ is defined in Sec.~\ref{lab:taskmodel}.

$hiTaskNum4Res(res, t)$ is the number of tasks competing for the $res$ with $t$ with higher priority defined as Eqn.~\ref{Eq_hiTaskNum4Res}. And $t' \gtrdot t$ indicates that the priority of 
$t'$ is higher than that of $t$.
% \[
% \small
% \begin{aligned}
% hiTaskNum4Res(t) := \# \left\{ t' \in \mathbb{T} \;\middle|\;
% \begin{array}{l}
% stage'(t') = stage'(t) \\
% \land\ t' \neq t \land t' \gtrdot t
% \end{array}
% \right\}
% \end{aligned}
% \]
% \[
% \small
% \begin{aligned}
% hiTaskNum4Res(res,t) := \# \left\{ t' \in \mathbb{T} \;\middle|\;
% \begin{array}{l}
% res \in M(t') 
% \land\ \\ t'  \neq t \land t' \gtrdot t
% \end{array}
% \right\}
% \end{aligned}
% \]
\begin{equation}
\small
\begin{aligned}
hiTaskNum4Res(res,t) := \# \left\{ t' \in \mathbb{T} \;\middle|\;
\begin{array}{l}
res \in M(t') 
\land\ \\ t'  \neq t \land t' \gtrdot t
\end{array}
\right\}
\end{aligned}
\label{Eq_hiTaskNum4Res}
\end{equation}

$nextStageStillUse(res)$ is the number of tasks still occupying resource $res$ at the next state, defined as follows.
\[
\begin{aligned}
                nextStageStillUse(res) := &\# \{ t' \in \mathbb{T} \mid (stage(t') = res) \\ 
                            & \ \ \ \ \ \land tick(t') > 1 \}
            \end{aligned}
\]

$stage'(t)$ represents the transition from the current stage to the next stage, and is formally defined as follows.
\begin{equation}
\small
stage'(t) := 
\begin{cases}
\texttt{Ready} & : stage(t) = \texttt{Block} \land \lnot resAble(t)\\
dispatch(t) & : stage(t) = \texttt{Block} \land resAble(t)\\
dispatch(t) & : stage(t) = \texttt{Ready}\\
\texttt{Finish} & : stage(t) \in \texttt{Exec}
\end{cases}
\label{Eq_stage_}
\end{equation}

$dispatch(t)$ assigns an instance of processing unit to $t$, depending on the scheduling algorithm, and returns the instance. 
% $alloc(t)$. 

$tick'(t)$ denotes the remaining time for the stage of $t$ after an transition of system, e.g., $udp$, formalized as follows.
\begin{equation}
tick'(t) := \max(0,\; tick(t) - 1)
\label{Eq_tick_}
\end{equation}

$ allocTime(t) $ denotes the execution time of task $t$ entering a new stage, as defined as follows.
\begin{equation}
 allocTime(t) = \max(\text{newStageTime}(t) - 1, 0)
 \label{Eq_altime}
\end{equation}

$newStageTime(t)$ is the execution time required for task $t$ when entering a new stage in an execution instance (schedule), as defined by Eqn.~\ref{Eq_newstageTime}. 
In Fig. 5, the same task may be assigned different execution times in two different executions. Task $t_1$ is assigned execution times of 40 and 30, respectively.
Specifically, except for the \texttt{Exec}, the execution time for other stages is 0, as they do not involve actual computation. The time for \texttt{Exec} called $allocESTime()$ is allocated from the range $onceESDTime(t, dev)$ on the assigned processing unit $dev$,  which is the interval $[BECT(t,dev), WCET(t,dev)]$. The scheduling policy determines 
resource allocation, and $onceESDTime(t, dev) $  may vary for $t$ across different $dev$. 
\begin{equation}
newStageTime(t) := 
\begin{cases}
\text{random from } [BE CT(t,dev), W\\ \ \ CET(t, dev)] : stage'(t) =\texttt{Exec} \\
0 \ \ \ \ \ \ \ \ \ \ \ \ \ \ \ \ \ :otherwise
\end{cases}
\label{Eq_newstageTime}
\end{equation}

Finally, starting from a system state $c$, the state after one transition is $upd(c)$, 
For example, as shown in Fig. 5, the system transitions from state $c_1$ to $c_2$ after one application of 
upd in an execution scenario. And after $n$ transitions of $c$, the state is $upd^n(c)$.

\section{Definition of Strict Timing Anomaly}
\begin{figure}
    \centering
    \includegraphics[width=1\linewidth]{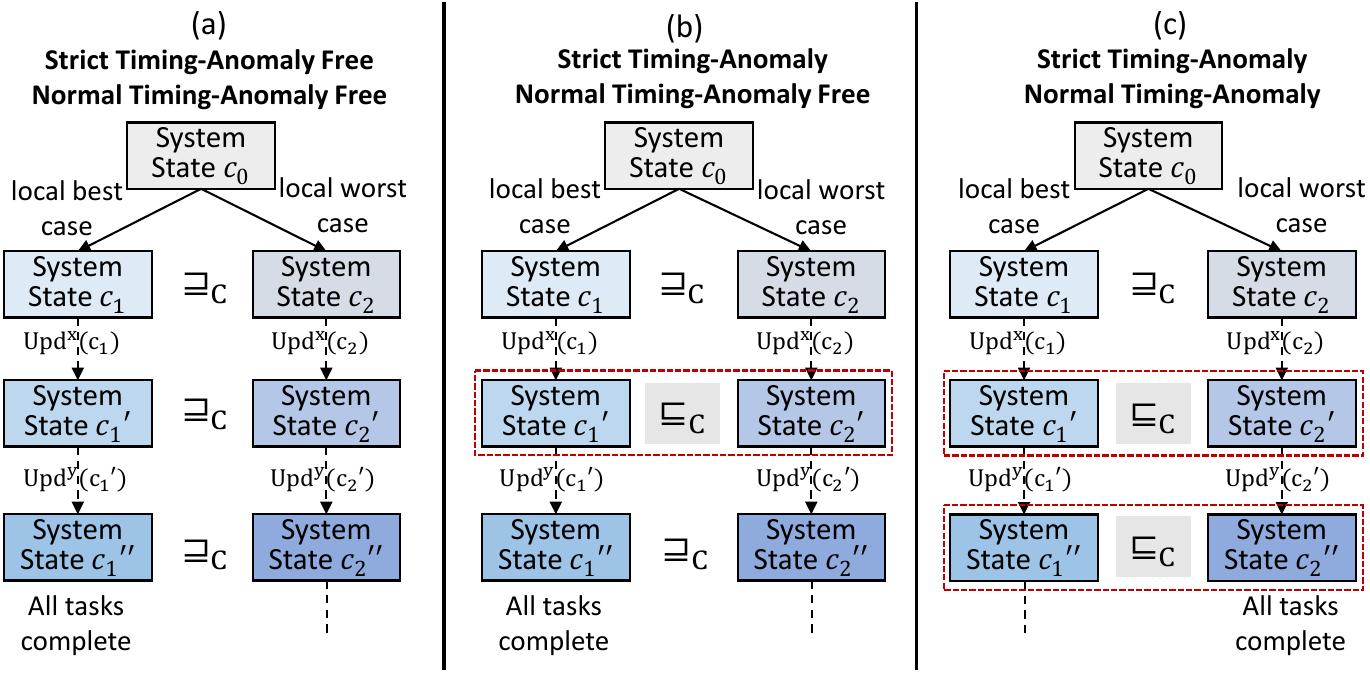}
    \vspace{-0.6cm}
    \caption{Strict timing anomalies and normal timing anomalies}
    \label{fig:TAdefine}
\end{figure}
We formalize the strict TA by extending the phenomenon of the normal TA, based on our execution progress models. Normal definitions~\cite{cassez2012timing, binder2021still, 9904748, 818824, reineke2006definition} focus only on whether an anomaly occurs in the system’s final completion state (i.e., the last task to finish), our strict definition strengthens the completion state of every task during the execution. Therefore, if a system is free of strict TA, it is guaranteed to be free of normal TA as well. Our definition is also general and, like recent advancements~\cite{binder2021still, 9904748}, supports multiple local reductions, rather than ~\cite{cassez2012timing, 818824, reineke2006definition} just a single reduction.

\textbf{Definition 1.} (Strict Timing Anomaly): Suppose the system evolves from the initial state to two different states \( c_1 \) and \( c_2 \) through valid execution paths. These two states satisfy the following condition, called \textbf{Condition 1}: 
% \begin{equation*}
% \begin{aligned}
% \exists \mathbb{S} \subseteq \mathbb{T}, \,  & (\forall t \in S,   \ c_2(t) \sqsubseteq_P c_3(t) )
%  \land (\forall t \in \mathbb{T} \setminus \mathbb{S}, c_2(t) = \\ c_3(t))  &  \land (allocFESTime(c_2, t) \geq allocFESTime(c_3, t)) 
% \end{aligned}
% \end{equation*}
\begin{equation*}
\begin{aligned}
& (\exists \mathbb{S} \subseteq \mathbb{T}, \,   \forall t \in \mathbb{S},   \ c_1(t) \sqsubset_P c_2(t) 
 \land \forall t \in \mathbb{T} \setminus \mathbb{S}, c_1(t) =  c_2(t))  \\ &  \land (\forall t \in \mathbb{T}, (c_1(t).stage () \sqsubset_S \texttt{Exec} \vee c_2(t).stage ()  \sqsubset_S \texttt{Exec})  \\ & \quad \quad \land  c_1(t).allocESTime() \geq c_2(t).allocESTime()) 
\end{aligned}
\end{equation*}

Then, the two execution state sequences starting from \( c_1 \) and \( c_2 \) are said to constitute \textit{a strict TA} if and only if:
\begin{equation*}
\exists n\in \mathbb{N}, \quad \lnot (upd^n(c_1) \sqsubseteq upd^n(c_2)).
\end{equation*}

$c_i(t).allocESTime()$ is the allocated execution time for the \texttt{Exec} stage of $t$ in an execution instance defined in Sec.~\ref{systemEST}.

The above definition can be interpreted as follows: For two system states  $c_1$ and $c_2$, if there exists a subset $\mathbb{S}$ of the system's task set $\mathbb{T}$, the execution progress of all tasks in $\mathbb{S}$ is smaller in state  $c_1$  than in $c_2$ (locally quicker), while the execution progress of tasks outside $\mathbb{S}$ remains the same in both $c_1$ and $c_2$. In state $c_1$ or $c_2$, for tasks in $\mathbb{T}$ that have not yet entered the \texttt{Exec} stage, their execution times for \texttt{Exec} stage allocated in the system evolved from state 
$c_1$ should be no less than those evolved from state $c_2$ (locally quicker).
Then a \textit{strict TA} occurs if, at some later time during execution, the state evolved from \( c_2 \) is not ahead of or not equal to the one evolved from \( c_1 \) in progresses, e.g., the Fig.~\ref{fig:TAdefine} (b) and (c).

Now, we reveal the connection between strict TA and all-WCETs based WCRT analysis. Consider  two system states \( c_1' \) and \( c_2' \), where \( \mathbb{S'} \subseteq \mathbb{S} \), and each task \( t' \in \mathbb{S'} \) has just entered the execution stage. In this case, \( c_1'.tick(t') \) is treated as the execution time of task \( t' \) in its current instance. So we can assume that it will execute using its WCET on the assigned processing unit. In contrast, we can assume that task $t'$ from \( c_2'\) will execute with a shorter time than its WCET. If the system progresses from \( c_1' \) and \( c_2' \) for the same duration, and at some point the system progress evolves from \( c_2' \) that is not ahead of or equal to that from \( c_1' \), then a \textit{strict TA} is said to occur. But, it doesn't mean that a \textit{normal} TA will occur, which only cares about the finish state of the system, e.g., the Fig.~\ref{fig:TAdefine} (b).

\textbf{Definition 2.} (Strict Timing Anomaly free): Conversely, if for any state pair $c_1$ and $c_2$ satisfying Conditions 1, the system always ensures that the progress of the evolution from $c_2$ is no less than that from $c_1$, that is, the system maintains monotonicity of execution
progress throughout its execution (Eqn.~\ref{noTAForm}), then the system is said to be \textbf{strict TA-free}.
\begin{equation}
\forall n\in \mathbb{N},\quad upd^n(c_1) \sqsubseteq upd^n(c_2)
\label{noTAForm}
\end{equation}

If the system in state $c_1$ assigns all-WCETs for tasks during execution, any other execution instances have greater progress, meaning 
others' response times are shorter. Clearly, if a system is strict TA-free, it is also normal TA-free, e.g., Fig.~\ref{fig:TAdefine} (a). Therefore, eliminating strict TAs can eliminate normal TAs.

\section{Constraints \& Proofs of Timing-Anomaly Free }
We introduce deterministic execution constraints as the foundation of TA-free dynamic scheduling. 
Then we prove its strict TA-free property, thereby avoiding normal TAs.

\subsection{TA-Free under Deterministic Execution Constraints}
\label{TA_free_DEC}
We argue that the cause of TA lies in the imbalance between the reduction from reduced task execution time during online execution may be insufficient to offset the increase in system execution time caused by a drop in system parallelism. The relationship between the two factors dynamically varies throughout execution, making precise analysis difficult.

To address this issue, we introduce deterministic execution constraints to ensure that the system exhibits \textit{monotonic execution progress}. Specifically, if a task has been completed in the offline simulated execution with all-WCETs, it must also be completed in the online execution; if a task is still in the execution stage in the offline schedule, its remaining execution time must be no less than that of the corresponding task in the online execution. Consequently, the response time of any task in the online execution is guaranteed to be no greater than its offline counterpart, meaning that the increase is always less than or equal to the reduction, thereby fundamentally eliminating TA. Therefore, the WCRT obtained from the all-WCETs execution offline is a safe value with our constraints.

In heterogeneous systems, where task execution time depends not only on workload variations but also the choice of processor units, the execution constraint should be applied to both execution order and resource allocation to eliminate TAs.

\textbf{Constraint 1.} (Resource Allocation Determinism): Tasks in the system must follow a predetermined resource allocation pattern \( resAllocPat_D \), where each pair \( (t_1, res_{t_1}) \) indicates that task \( t_1 \) is only allowed to execute on processing units with type \( res_1 \) during online execution. This constraint prevents \( t_1 \)
from selecting other types of processing units with different execution speeds, which would result in varying WCETs.
\begin{equation*}
resAllocPat_D = \{(t_1, res_{t_1}) ,(t_2, res_{t_2}),...,(t_n, res_{t_n})\}
\end{equation*}

\textbf{Constraint 2.} (Execution Order Determinism): Tasks must enter the execution stage according to an offline legal schedule order \( O_D \). The relation \( \preceq \) denotes the execution order between tasks, where \( t_1 \preceq t_2 \) means that task \( t_1 \) must begin execution no later than task \( t_2 \), i.e., \( t_1 \) has a higher or equal priority in the execution sequence. A legal order implies that the execution order respects all task dependencies in the DAG.
\begin{equation*}
O_D = (t_1 \preceq t_2 \preceq  t_3 ... \preceq t_N)
\end{equation*}

\subsection{Formal Proof of TA Free with the Constraints}
Here, we prove that if systems execute with our constraints, they will not exhibit strict TA, i.e., prove Definition 2 holds. 

We begin by proving its special case Lemma 4, where the task set \( \mathbb{S} \) contains only a task, e.g., a local reduction. Based on this, we generalize the result to the case where \( \mathbb{S} \) is any finite set of tasks, e.g.,  Theorem 3. This completes a proof of the system's strict TA-free and, therefore, its normal TA-free.

Before proving Lemma 4, we first prove that the system's state evolution exhibits forward progress, e.g., Lemma 1, and monotonicity, e.g., Lemma 2 and 3. The proofs of Lemmas 1, 2, and 3 are provided in the appendix.

\textbf{Lemma 1.} (Forward Progress Property)
% The system's execution progress strictly increases after each state transition.
The system's execution progress strictly increases after each state transition, if there are tasks that have not reached the \texttt{Finish} stage, i.e.,
\begin{equation*}
\forall c\in C: \exists t \in \mathbb{T}, c.stage(t) \neq \texttt{Finish} \Rightarrow c\sqsubset_{C} udp(c)
\end{equation*}

Lemma 1 implies $\forall t \in \mathbb{T},upd^n(c_2)(t) \sqsubseteq_P upd^{n+1}(c_2)(t)$. But at least one task $t'$ has $upd^n(c_2)(t') \sqsubset_P upd^{n+1}(c_2)(t')$.

\textbf{Lemma 2.} (Monotonicity of Dependency Removal) 
Let $c_1, c_2 \in C$ be two valid states with $c_1 \sqsubseteq c_2$, it holds that:
\begin{equation*}
\forall t \in \mathbb{T}: \quad c_1.depComp(t) \Rightarrow c_2.depComp(t)
\end{equation*}

\textbf{Lemma 3.} (Monotonicity of Execution Stage)  Let \( c_1, c_2 \in C \) be two valid system state satisfying \( c_1 \sqsubseteq c_2 \), it holds that:
\begin{align*}
\forall t \in \mathbb{T}: \quad & c_1.stage (t) \sqsubset_S upd(c_1).stage(t) \Rightarrow \\
& \quad \quad \quad upd(c_1).stage(t) \sqsubseteq_S upd(c_2).stage(t)
\end{align*}

 Now, we use the above lemmas to prove the execution progress of the system is monotonicity with our constraints which means a strict TA-free property , e.g., Theorem 3.

\textbf{Theorem 3.}  
Under constraints 1 and 2, the system is free of strict TA; i.e., Definition 2 holds under these constraints.

To prove Theorem 3, we first consider its special case:

\textbf{Lemma 4.}  
When the $\mathbb{S}$ satisfies $|\mathbb{S}| = 1$, Theorem 3 holds. 

\begin{proof}
We prove by mathematical induction.

\textbf{Base Case:}  
When the system transition function $upd$ is applied $n = 0$, it clearly satisfies \( upd^n(c_1) \sqsubseteq upd^n(c_2) \).

\textbf{Inductive Step:}  
Assume \( n > 0 \), the \( upd^n(c_1) \sqsubseteq upd^n(c_2) \) holds.  
We now prove that \( upd^{n+1}(c_1) \sqsubseteq upd^{n+1}(c_2) \)  holds. From task perspective, it means \( \forall t \in \mathbb{T}, upd^{n+1}(c_1)(t) \sqsubseteq_P upd^{n+1}(c_2)(t) \)  holds. Proved by the following cases:

\begin{itemize}
    \item \textbf{Case 1:} $upd^n(c_1)(t) = upd^{n+1}(c_1)(t)$.  In this case, task $t$ remains in the last state. Thus, we have $ upd^n(c_1)(t) = upd^{n+1}(c_1)(t) \sqsubseteq_P upd^n(c_2)(t)$. And by Lemma 1, we have $upd^n(c_2)(t) \sqsubseteq_P upd^{n+1}(c_2)(t)$, thus $upd^{n+1}(c_1)(t) \sqsubseteq_P upd^{n+1}(c_2)(t)$.

    \item \textbf{Case 2:} $upd^n(c_1).stage(t) = upd^{n+1}(c_1).stage(t)$ and $upd^{n+1}(c_1).tick(t) = upd^n(c_1).tick(t) - 1$.  
    That is the case where task $t$ in $c_1$ state remains in the stage \texttt{Exec}, but the remaining time decreases in this stage.  
    We further divide this case into sub-cases by $udp^n$:
    \begin{itemize}
        \item If $upd^n(c_1)(t) = upd^n(c_2)(t)$, by the definition of $upd()$ and the assumption for this case, the only change is a decrease of 1 in their remaining time. Therefor, we have $upd^{n+1}(c_1)(t) = upd^{n+1}(c_2)(t)$.
        \item If $upd^n(c_1)(t) \sqsubset_P upd^n(c_2)(t)$, we further divide the sub-case into the following cases by the stage:
        \begin{itemize}
            \item If $upd^n(c_1).stage(t) \sqsubset_S upd^n(c_2).stage(t)$, it follows $upd^n(c_2).stage(t)$ is \texttt{Finish}, and since $upd^n(c_2).stage(t)$ and $upd^{n+1}(c_2).stage(t)$ are both \texttt{Exec} stage. Therefore, it implys that $upd^{n+1}(c_1)(t) \sqsubset_P upd^{n+1}(c_2)(t)$ holds.
            \item If $upd^n(c_1).stage(t) = upd^n(c_2).stage(t)$, we have $upd^n(c_1).tick(t) > upd^n(c_2).tick(t)$.
            \begin{itemize}
                \item If $upd^n(c_2).tick(t) > 1$, task $t$'s execution times in $upd^n(c_1)$ and $upd^n(c_2)$  decrease by 1 for $tick'(t)$ after a $udp$. Therefore, we have $upd^{n+1}(c_1).tick(t) > upd^{n+1}(c_2).tick(t) \geq 1$, implying $upd^{n+1}(c_1)(t) \sqsubset_P upd^{n+1}(c_2)(t)$.
                \item If $upd^n(c_2).tick(t) = 1$, then in the next state $c_2$ moves to the \texttt{Finish} stage while $c_1$ remains in the \texttt{Exec} stage, leading to $upd^{n+1}(c_1)(t) \sqsubset_P upd^{n+1}(c_2)(t)$.
            \end{itemize}
        \end{itemize}
    \end{itemize}

    \item \textbf{Case 3:} $upd^n(c_1).stage(t) \sqsubset_S upd^{n+1}(c_1).stage(t)$.  
    That is, task $t$ moves to a new stage in the state $upd^{n+1}(c_1)$.  
    % By Lemma 3 we have $upd^{n+1}(c_1).stage(t) \sqsubseteq_S upd^{n+1}(c_2).stage(t)$. 
    We divide this case into sub-cases based on whether $upd^{n+1}(c_1).stage(t)$ is the \texttt{Exec}:
    \begin{itemize}
    \item If $upd^{n+1}(c_1).stage(t) \neq \texttt{Exec}$, only need to compare the stages. If the stage of $c_1$ changes, we can get $upd^{n+1}(c_1).stage(t)$ $\sqsubseteq_S$ $upd^{n+1}(c_2).stage(t)$ by Lemma 3. And if the stage of $c_1$ does not change, it clearly also holds by Lemma 1. So the $upd^{n+1}(c_1)(t)$ $\sqsubseteq_{P}$ $ upd^{n+1}(c_2)(t)$ holds.
    \item If $upd^{n+1}(c_1).stage(t) = \texttt{Exec}$, we consider the following two cases:
        \begin{itemize}
        \item If $upd^{n+1}(c_1).stage(t) \neq upd^{n+1}(c_2).stage(t)$, we can also directly compare the stages, similar to the previous case.
        \item If $upd^{n+1}(c_1).stage(t) = upd^{n+1}(c_2).stage(t)$, they both are in the \texttt{Exec}, so we need to compare their remaining execution times.
        One case is that task $t$ enters \texttt{Exec} simultaneously in $upd^{n+1}(c_1)$ and $upd^{n+1}(c_2)$. By the conditions of Definition 2, the $allocESTime()$ of $t$ in the $c_1$ case must be $\geq$ to that in the $c_2$ which is also the remaining execution time. Thus, it holds $upd^{n+1}(c_1)(t) \sqsubseteq_{P}$ $ upd^{n+1}(c_2)(t)$. The validity of this condition relies on the relationship of two $allocESTime()$, which depends on the processor selected for the task in both cases. Constraint 2 ensures that the processors are chosen with the same WCET from the given $ resAllocPat_D$ in both cases to satisfy the condition. The boundary case occurs when task $t$ executes by the same WCET in both cases. 
        The second case is that the task $t$ enters \texttt{Exec} before $upd^{n+1}(c_2)$. Clearly, under the Constraint 2, we also have $upd^{n+1}(c_1)(t) \sqsubset_{P}$ $ upd^{n+1}(c_2)(t)$.
        \end{itemize}
    \end{itemize}
    % PPPPPPPPPPPPPPPPPPPPPPPPPPPPPPPPPPPP
    % Note that $t$ cannot be a task in $\mathbb{S}$, because tasks in $\mathbb{S}$ are already in the \texttt{Exec} stage. When it enters this stage, its execution duration is $stageTime(t)$. According to the condition from Definition 2, the duration is equal between $c_1$ and $c_2$. To ensure this condition holds, we must have Constraint 1 to guarantee that task $t$ is assigned to identical-speed processing units in states $c_1$ and $c_2$. Otherwise, differences in speeds between units (e.g., $execTimeOnDev(t, dev_1) \neq execTimeOnDev(t, dev_2)$) could result in inconsistent durations for $c_1$ and $c_2$, invalidating the condition.

    % Therefore, if task $t$ enters the \texttt{Exec} stage in the next state, then $upd^{n+1}(c_1).tick(t) = time' \geq upd^{n+1}(c_2).tick(t)$, because task $t$ is also at least in the \texttt{Exec} stage in $upd^{n+1}(c_2)$. If task $t$ transitions to a non-execution stage, then we have $upd^{n+1}(c_1)(t) = upd^{n+1}(c_2)(t)$. Thus, we conclude that $upd^{n+1}(c_1) \sqsubseteq upd^{n+1}(c_2)$.

    Hence, in all cases, we have: $upd^{n+1}(c_1) \sqsubseteq upd^{n+1}(c_2).$
\end{itemize}
Thus, Lemma 4 is proved.
\end{proof}

We now prove the general case of Theorem 3. 

\begin{proof}
We also proceed by mathematical induction. By constructing two intermediate states $c_3$ and $c_4$. And $c_3$ differs from $c_1$ in the execution progress of $m$ tasks, $c_4$ differs from $c_3$ in one task, and $c_4$ differs from $c_2$ in $m$ tasks. By the base case, which is Lemma 4 (one task with different progress) and the induction hypothesis ($m$ tasks with different progress), we establish the desired result for $m+1$ tasks. Specifically: 

\textbf{Base case}: $|\mathbb{S}| = 1$ corresponds to Lemma 4 proved. 

% \textbf{Inductive Step:}  We now assume the conclusion holds for $|\mathbb{S}| = m$ and aim to prove it for $|\mathbb{S}| = m + 1$. From defination 2, it has $\forall t \in \mathbb{S},\; c_1(t) \sqsubset_{P} c_2(t)$ and $\forall t \in \mathbb{T} \setminus \mathbb{S}, c_1(t) =  c_2(t)$. Now, we extract a task $t' \in \mathbb{S}$, and $c_1(t') \sqsubset_{P} c_2(t')$ holds. Let $ x_1 = c_1.\text{tick}(t'), X_1 = c_1.\text{stage}(t')$ and $ y_2= c_2.\text{tick}(t'), Y_2 = c_2.\text{stage}(t')$. And we have $(X_1 \sqsubset_{S} Y_2) \vee (X_1 = Y_2 \land x_1 > y_2)$.

\textbf{Inductive Step:}  We now assume the conclusion holds for $|\mathbb{S}| = m$ and aim to prove it for $|\mathbb{S}| = m + 1$. From defination 2, it has $\forall t \in \mathbb{S},\; c_1(t) \sqsubset_{P} c_2(t)$ and $\forall t \in \mathbb{T} \setminus \mathbb{S}, c_1(t) =  c_2(t)$. Now, we extract a task $t' \in \mathbb{S}$, and $c_1(t') \sqsubset_{P} c_2(t')$ holds. Let $ x = c_1.tick(t'), X = c_1.stage(t')$ and $ y= c_2.tick(t'), Y = c_2.stage(t')$. And we have $(X \sqsubset_{S} Y) \vee (X = Y \land x > y)$.

First, we construct an intermediate state $c_3$ satisfying that: $(\forall k \in \mathbb{S} \setminus \{t'\}, c_1(k)\sqsubset_{P} c_3(k) \sqsubset_{P} c_2(k)) \land (c_3.tick(t') = x \land c_3.stage(t') = X) \land ( \forall k \in \mathbb{T} \setminus \mathbb{S},\; c_3(j) = c_1(j)) $. Since $c_1$ and $c_3$ differ only on $m$ tasks, and the execution progress of the $m$ tasks in $c_1$ is strictly less than that in $c_3$. By the induction hypothesis for $|\mathbb{S}|=m$, we have: $upd^n(c_1) \sqsubseteq upd^n(c_3).$

Second, we construct another intermediate state $c_4$ such that: $(\forall k \in \mathbb{S} \setminus \{t'\}, c_4(k) = c_3(k)) \land (c_4.tick(t') = y \land c_4.stage(t') = Y) \land ( \forall j \in \mathbb{T} \setminus \mathbb{S},\; c_4(j) = c_3(j)) $. Therefor, $c_3$ and $c_4$ differ in only task $t'$, meanwhile $c_3(t') \sqsubset_{P} c_4(t')$. So, according to the base case, we have: $upd^n(c_3) \sqsubseteq upd^n(c_4)$.

Third, we can observe that $c_4$ and $c_2$ satisfy: $(\forall k \in \mathbb{S} \setminus \{t'\}, c_4(k) \sqsubset_{P} c_2(k)) \land ( \forall j \in (\mathbb{T} \setminus \mathbb{S}) \cup {t'},\; c_4(j) = c_2(j))$. So, $c_2$ and $c_4$ differ only on $m$ tasks, and the execution progress of these tasks in $c_4$ is strictly less than that in $c_2$, which again satisfies the induction hypothesis, so $upd^n(c_4) \sqsubseteq upd^n(c_2)$.

Last, we conclude the following chain of monotonic relations: $ upd^n(c_1) \sqsubseteq upd^n(c_3) \sqsubseteq upd^n(c_4) \sqsubseteq upd^n(c_2)$, implying $upd^n(c_1) \sqsubseteq upd^n(c_2)$ holds.
\end{proof}

In conclusion, under our execution constraint, Definition 2 is satisfied, implying the system is monotonic and thus strict TA-free. Consequently, the system is also free of normal TAs.

\section{Implementation of Deterministic Dynamic Execution and Constraint Strategy Design}

This chapter first gives two methods for deriving the execution constraints defined in Sec.~\ref {TA_free_DEC} and then presents a TA-free dynamic scheduling algorithm, termed Deterministic Dynamic Execution (DDE), which is based on the constraints.

\subsection{Two Methods for Deriving Execution Constraints}
Before implementing DDE, it is necessary to derive the deterministic execution constraints offline, including the execution order and resource allocation pattern. However, different constraints can significantly affect the WCRT. Here, we give two methods to derive the constraints.

% \begin{algorithm}[t]
% \caption{HJACP Algorithm for Execution Constraints }
% \label{alg:heuristic-jitter}
% \textbf{Input:} DAG $G = (V, E)$, execution time $\tau_{\text{WCET}}$, $\tau_{\text{BCET}}$, resource availability $A$, resource counts $C$, weight $\beta \in [0, 1]$ \\
% \textbf{Output:} Execution Constraint $EC$, $WCRT$

% \begin{algorithmic}[1]
% \State Initialize $\text{rank}[v] \gets 0$, $S \gets \emptyset$
% \For{each task $v \in V$}
%     \State Compute $\text{rank}[v]$ using average WCET weighted by $C$ and critical successor cost
% \EndFor
% \State Sort $V$ in descending order of rank to get list $T$

% \For{each task $v \in T$}
%     \State $bestCost \gets \infty$, $bestRes \gets \bot$
%     \For{each resource $r$ supporting $v$}
%         \State $s_W \gets \max(A^{\text{W}}[r], \max_{u \in pred(v)} S[u].\text{finish}_W)$
%         \State $s_B \gets \max(A^{\text{B}}[r], \max_{u \in pred(v)} S[u].\text{finish}_B)$
%         \State $f_W,f_B \gets s_W + \tau_{\text{WCET}}[v][r],s_B + \tau_{\text{BCET}}[v][r]$
%         \State $cost \gets \beta \cdot f_W + (1 - \beta) \cdot (f_W - f_B)$
%         \If{$cost < bestCost$}
%             \State Update $bestCost, bestRes, s^*_W\gets cost, r, s_B$
%         \EndIf
%     \EndFor
%     \State Schedule $v$ on $bestRes$, update $S[v]$
%     \State Update $A^{\text{W}}[bestRes], A^{\text{B}}[bestRes] \gets f_W, f_B$
% \EndFor
% \State $WCRT \gets \max_v S[v].\text{finish}_W$
% \State Extraction of Execution Constraint to $EC$
% \State \textbf{return} EC, WCRT
% \end{algorithmic}
% \end{algorithm}

\begin{algorithm}[t]
\caption{HACPA Algorithm for Execution Constraints}
\label{alg:heft-jitter-recursive}
\textbf{Input:} DAG $G = (V, E)$, execution time $\tau_{\text{WCET}}$, available processor units number $\mathbb{P}$ \\
\textbf{Output:} Execution Constraint $EC$, $WCRT$

\begin{algorithmic}[1]
\State Initialize $\text{rank}[t],S[t] \gets 0,\emptyset$ for all tasks $t \in \mathbb{T}$

\Function{calc\_rank}{task $t$}
    \If{$\text{rank}[t] > 0$} 
        \State \textbf{return} $\text{rank}[t]$ // Return pre-computed rank
    \EndIf
    \State $succ\_t \gets \text{get successors of task } t$
    \State $avg\_cost\_t \gets$ the average cost on different types of processor units for task $t$
    \If{$succ\_t$ is empty} 
        \State $\text{rank}[t] \gets avg\_cost\_t$ // WCET for task $t$
    \Else
        \State $\text{rank}[t] \gets avg\_cost\_t + \max_{s \in succ\_t} \text{calc\_rank}(s)$
    \EndIf
    \State \textbf{return} $\text{rank}[t]$
\EndFunction

\For{each task $t \in \mathbb{T}$}
    \State \text{calc\_rank}(t) // Calculate rank recursively
\EndFor

\State Sort $\mathbb{T}$ in descending order of $\text{rank}$ for sorted task list $\mathbb{L}$
\State Initialize $avil\_time[res] \gets 0$ for all $res \in \mathbb{P}$
\For{each task $t \in \mathbb{L}$}
    \State Initialize $bestRes, bestS, bestF \gets \bot, \infty, \infty$
    \For{each resource $res$ supporting task $t$}
        \State $s \gets \max({avil\_time}[res], \max_{u \in pred(t)} S[u].f)$
        \State $f \gets s + \tau_{\text{WCET}}[t][res]$ // Compute finish times
        \If{$f < bestF$}
            \State $bestS, bestF, bestRes,  \gets s, f, res$
        \EndIf
    \EndFor
    \State $S[t].s, S[t].f, res[t] \gets bestS, bestF, bestRes$
    \State $avil\_time[res] \gets bestF$
\EndFor
\State $WCRT \gets \max_{t \in \mathbb{T}} S[t].f$
\State Extract Execution Constraint $EC$ from $S$
\State \textbf{return} EC, $WCRT$
\end{algorithmic}
\label{alg:1}
\end{algorithm}

\subsubsection{Method 1: from the Trace of all-WCETs Execution}
\ 
\newline
\indent
\label{ger_cons_method1}
First, a fast and straightforward method based on the execution trace. We can extract the constraints from the execution trace, which is from the original dynamic scheduling algorithm like HBFS under the all-WCETs simulated execution. This trace records, for each task, its start and finish execution times as well as the assigned processing unit.

Based on this trace, we construct the execution order $O_D$ for the constraint by imposing a total order on tasks according to their start execution times, with earlier-starting tasks given higher rank; If tasks have the same start time, the task with the smaller ID is given higher rank.

For the resource allocation pattern $resAllocPat_D$, we derive each task $t'$ eligible resource set $res_t$ from the processing unit to which it is assigned in the trace. Specifically, all processing units with the same execution speed—equivalently, these processor units with the same type and this group is taken as the allocatable resource set $res_t$ for task $t$.

For example, we can get the constraints from the Fig.~\ref{fig:twoExectionCase}(b) for the DAG in Fig.~\ref{fig:DAG1} under HFCFS. The execution order $O_D$ is: $t_1 \preceq t_2 \preceq t_3$ $ \preceq t_6$ $ \preceq t_4$ $ \preceq t_5 \preceq t_7$. And the resource allocation $resAllocPat_D$ is: $(t_1, CPU_0)$, $(t_2, CPU_0)$, $(t_3, CPU_1)$, $(t_4, GPU_1)$, $(t_5, GPU_1)$, $(t_6, GPU_0)$, $(t_7, CPU_0)$.

\subsubsection{Method 2: from a Heuristic Algorithm}
\ 
\newline
\indent
% To minimize the system's WCRT, we design a heuristic algorithm that generates execution constraints with WCRT reduction as the optimization goal. The algorithm is heterogeneous-aware and critical-path-aware, along with scheduling decisions aimed at minimizing the WCRT. The critical path refers to the longest path in the DAG from the source node to the sink node, while considering that each task may choose different computing resources.
To minimize the system's WCRT, we design a heuristic algorithm called HACPA, as shown in Alg.~\ref{alg:1}, which generates execution constraints with the goal of reducing WCRT. The algorithm is both heterogeneous-aware and critical-path-aware, along with scheduling decisions aimed at minimizing the WCRT. The critical path refers to the longest path in the DAG from the source to the sink node, considering that each task may choose different computing resources.

The algorithm first assigns a rank to each task. As shown in Alg.~\ref{alg:1}, the rank is defined as the weighted average of the task’s WCETs across different types of processor units in line 7 of Alg.~\ref{alg:1}, combined with the maximum rank among its successor tasks ($succ\_t$). The rank is computed recursively in a bottom-up manner, propagating critical path costs from the leaf to the source node. This strategy fully leverages the global information of the DAG and effectively identifies critical tasks that dominate the system’s WCRT.

Next, tasks are assigned to processing units in descending order of their rank. 
For each task, a processing unit is selected and executed with its corresponding WCET. The resource selection considers the unit's available time, the task's precedence dependencies ($pred(t)$), and its WCET to minimize the task's WCRT. Meanwhile, the resource allocation is recorded in the $res$, along with the corresponding task start and end times recorded in the $S$, which together form the trace.

Finally, once the trace is obtained, the process of extracting execution constraints follows the same steps as in Method 1.

\begin{figure}
    \centering
    \includegraphics[width=1\linewidth]{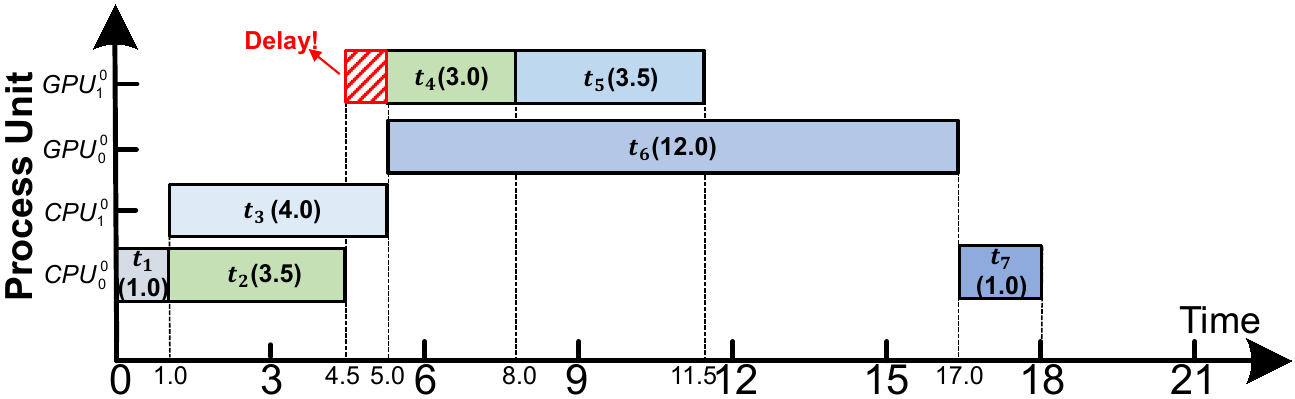}
    \caption{A case for deterministic dynamic execution to eliminate TA in Fig.~\ref{fig:TA-case}}
    \label{fig:DeterministicTA}
\end{figure}

\subsection{Implementation for Deterministic Dynamic Execution}
We employ a delayed execution mechanism to implement DDE based on our constraints.
Specifically, during online execution, when a task is selected from the ready queue, it is not necessarily assigned computing resources or started immediately. The system checks whether there exist any predecessor tasks—according to the offline given execution order $O_D$—that have not yet begun execution. If all such predecessors have started, the task is then scheduled to execute on one of the available processing units, with the type specified by the offline given resource allocation pattern $resAllocPat_D$. Otherwise, the task's resource allocation and execution are delayed until all preceding tasks in the $O_D$ have started.

Obtain the WCRT for a system under DDE. We note that, according to Definition 2, the system's execution progress is monotonic under our execution constraints, ensuring the absence of TAs. Therefore, the response time obtained from all-WCETs execution represents the system's WCRT. The two methods for deriving the constraints, based on all-WCETs, yield the system's WCRT as the resulting response time.

Here, we provide an example of a system using DDE, based on the DAG shown in Fig.~\ref{fig:DAG1}. We directly apply Method 1 to extract execution constraints, using the corresponding all-WCETs execution trace shown in Fig.~\ref{fig:twoExectionCase}(b). These constraints were previously presented in Sec. \ref{ger_cons_method1}. As a result, the system's WCRT is 18, corresponding to the response time in Fig.~\ref{fig:twoExectionCase}(b). This process also eliminates TAs, as detailed below:

As shown in Fig.~\ref{fig:DeterministicTA}, the TA in Fig.~\ref{fig:TA-case} is eliminated under the above constraints. Specifically, after task $t_2$ finishes, task $t_4$ could have been scheduled on $GPU_0^0$, but due to the constraint on execution order $O_D$, it must wait until $t_6$ starts executing before $t_4$ is allowed to begin. Moreover, $t_4$ is constrained to execute only on instances of processor units type $GPU_1$ from the given resource allocation pattern  $resAllocPat_D$. As a result, its execution is delayed until time 5.

It is worth noting that while the above example appears to fix tasks to specific instances of processing units, seemingly static, this is only due to the system having only one instance per type of processing unit. Our DDE inherently supports dynamic scheduling among instances with identical execution speeds of the same type, e.g., with the same WCET.

\section{EVALUATION}
% We first verify that deterministic dynamic execution can effectively eliminate TA by experiments. Then we analyze the effectiveness of deterministic dynamic execution and the HACPA algorithm to get better execution contains on randomly generated DAGs, evaluating metrics including WCRT, jitter of response time, and average response time.

This chapter validates the effectiveness of our method. First, we verify that deterministic dynamic execution (DDE) can eliminate TAs in a system with timing anomalies (TAs). Then, using a large set of randomly generated systems, we also verify that it can eliminate TAs and compare the multiple metrics of DDE under two constraint generation methods, with traditional dynamic scheduling algorithms. The metrics include WCRT, response time jitter, and average response time.

% \subsection{A Case for Verification of Deterministic Execution for Eliminating TA}

\subsection{Verification with a Case for Eliminating TA}
\begin{figure}[t]
    \centering
    \includegraphics[width=1\linewidth]{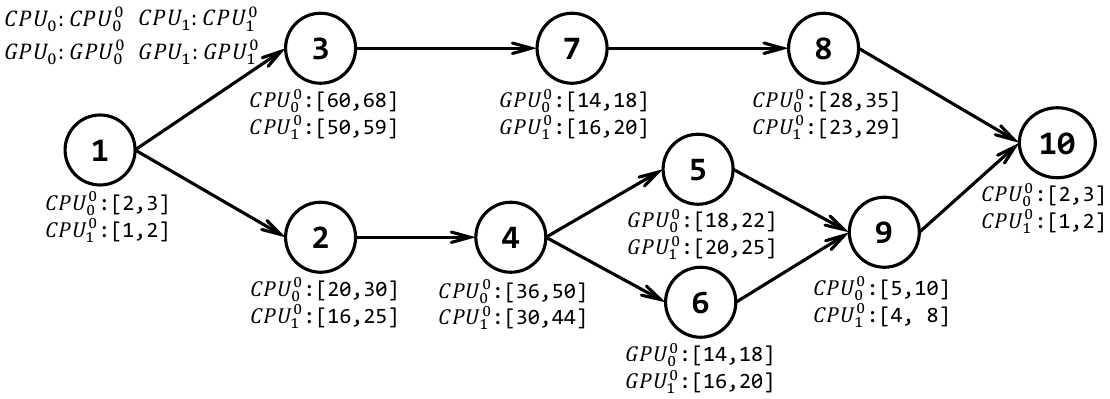}
    \caption{A multi-typed DAG with TA under HFCFS Scheduling}
    \label{fig:DAG2}
\end{figure}

This section uses a system exhibiting TAs to demonstrate that DDE can not only eliminate TAs but also reduce the system's WCRT. As shown in Fig.~\ref{fig:DAG2}, the system is represented by the DAG consisting of 10 nodes, and the compute resource and corresponding interval of execution times are also shown.
% \begin{figure}[t]
%     \centering
%     \includegraphics[width=0.9\linewidth]{实验/boxplot_with_custom_xticks_larger_font.pdf}
%     \caption{Execution time distribution of a DAG system with TA}
%     \label{fig:Adag_TA_execution_time}
% \end{figure}

\begin{figure}[t]
    \centering
    \includegraphics[width=0.9\linewidth]{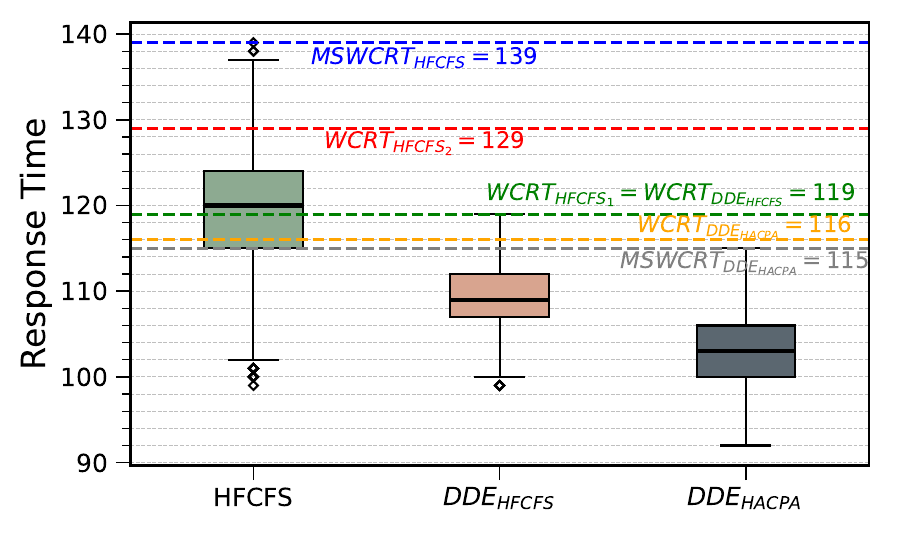}
    % \vspace{-0.6cm}
    \caption{Execution Time Distribution of the DAG in Fig.~\ref{fig:DAG2} under HFCFS and two DDE Scheduling Algorithms ($DDE_{HFCFS}$ and $DDE_{HACPA}$)}
    % \vspace{-0.4cm}
    \label{fig:Adag_TA_execution_time}
\end{figure}

In this experiment, 10000 online executions with various execution times from its interval of BCET and WCET are performed under the HFCFS algorithm, which preferentially schedules the first-ready tasks and prioritizes the processors with fast processing speeds. The distribution of its online execution time is the first box plot in Fig.~\ref{fig:Adag_TA_execution_time}. 
From the all-WCETs offline simulation of HFCFS, the corresponding $WCRT_{HFCFS_1}$ is 119. And a more conservative offline bound, $WCRT_{HFCFS_2}=129$, is obtained by the method described in Sec.~\ref{mov}. 
The observed maximum online execution time, $MSWCRT_{HFCFS} = 139$ exceeds $WCRT_{HFCFS_1}$ and $WCRT_{HFCFS_2}$, confirming the presence of TA.

% To eliminate TA in this system, we apply the DDE. 
We apply the DDE to eliminate TA in this system with the constraints of 
execution order $O_D $ is $ t_1 \preceq t_2 \preceq t_3 \preceq t_4 $ $\preceq t_7 \preceq v_5 $ $\preceq t_6 \preceq t_8 \preceq t_9 \preceq t_{10}$ and the resource allocation pattern $resAllocPat_D$ is $(t_1, CPU_0)$, $(t_2, CPU_1)$, $(t_3, CPU_0)$, $(t_4, CPU_1)$, $(t_5, GPU_1)$, $(t_6, GPU_0)$, $(t_7, GPU_0)$, $(t_8, CPU_1)$, $(t_9, CPU_0)$, $(t_{10}, CPU_1)$. These constraints are derived from the scheduling trace obtained from the all-WCETs simulated execution that produced $WCRT_{HFCFS_1}$. An enumeration of online executions with varying execution times is performed for DDE. The result is the orange box in Fig.~\ref{fig:Adag_TA_execution_time} with label $DDE_{HFCFS}$, which indicates the execution time never exceeds its WCRT, e.g., $WCRT_{DDE_{HFCFS}}$. Therefore, under DDE, there are no TAs.

From the box plots corresponding to the HFCFS algorithm, the measured worst-case response time $MSWCRT_{HFCFS}$ can be observed as 139, although this does not exclude the possibility that there are worse executions that have not yet been captured. 
In contrast, after adopting our DDE under the constraints from itself, the WCRT is $WCRT_{DDE_{HFCFS}} = 119$. This shows that DDE not only eliminates TA but also can reduce the WCRT of the system by at least: $\frac{139 - 119}{139} \times 100\% = 14.4\%$. Given that 139 may not be the true worst-case scenario for the system without constraints, the performance improvement may be even more significant. We can also find the response time jitter is also reduced from the distribution.

In addition, using our HACPA algorithm to derive new execution constraints for DDE, we can get a smaller WCRT, i.e., $WCRT_{DDE_{HACPA}}= 116$.

\begin{figure*}[t]
    \centering
    \includegraphics[width=\textwidth]{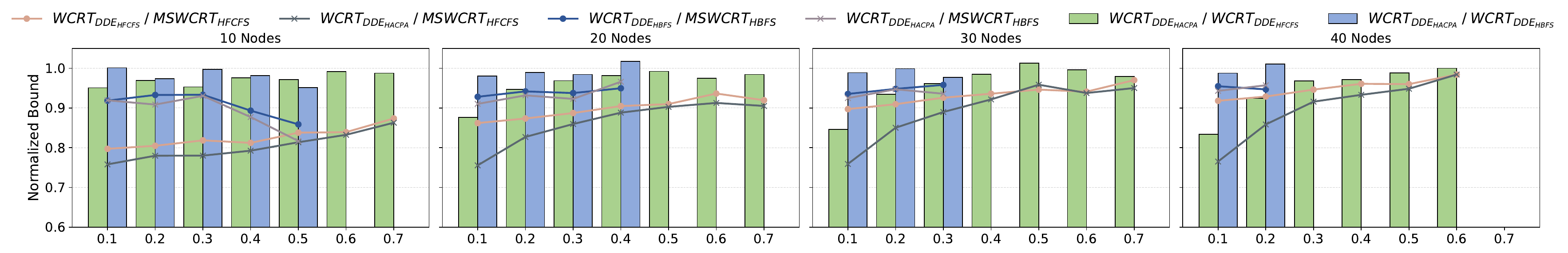}  % 替换为你的图片文件名
    \caption{$WCRT$ and $MSWCRT$ for multi-type DAGs with various configurations of DAGs under several different scheduling algorithms}
    \label{muliteDAGcpm}
\end{figure*}

% \begin{figure*}[t]
%     \centering
%     \includegraphics[width=\textwidth]{实验/longpictureExper333.pdf}  % 替换为你的图片文件名
%     \caption{Comparison of $WCRT$ and $MSWCRT$ for multiple DAGs with different numbers of nodes under several different scheduling algorithms}
%     \label{muliteDAGcpm}
% \end{figure*}
% \subsection{Validation of Deterministic Execution Using Randomly Generated DAGs}

\subsection{Validation with Randomly Generated DAGs}
This section evaluates the WCRT, response time jitter, and average response time of DDE with the two constraint derivation methods, using randomly generated multi-typed DAGs. 

Here, we compare the normal dynamic scheduling algorithms HFCFS and HBFS with our TA-free dynamic scheduling algorithm DDE. The constraints for DDE are derived from all-WCETs simulations under HFCFS, HBFS, and our heuristic algorithm HACPA, resulting in the corresponding DDE algorithms: $DDE_{HFCFS}$, $DDE_{HBFS}$, and $DDE_{HACPA}$. 

The DAGs are generated using the $G(n, p)$ model provided by the RDGEN~\cite{RDGEN} tool. It is important to note that this tool can only generate standard non-type DAGs and does not support the generation of multi-typed DAGs for heterogeneous systems. Therefore, in this work, we only utilize the DAG topology generated by RDGEN and independently assign node types and execution times to the corresponding processor units. 
When converting the DAGs to multi-typed DAGs, 100 different combinations of processing unit types and corresponding execution times were assigned to each DAG.
The generation involves the following parameters: The number of nodes in a DAG is $(10, N_{max})$, where $N_{max}$ is 40. The interconnection probability of the nodes is $p$, which denotes the probability that there is a connected edge between two nodes in the DAG, $p \in \{0.1, 0.2,... ,0.9\}$. Except for the source and sink nodes, which are fixed to run on CPUs, the remaining nodes are randomly assigned to all processing units, reflecting the typical execution pattern of heterogeneous applications. The execution time of a node in the corresponding execution unit is $[C_{min}, c_{max}]$, where $C_{min} = rand[1, 1000]$, $C_{max} = rand(x_1, x_2) \times C_{min}$ there are $80\%$ of nodes use $x_1 = 10$, $x_2 = 30$ and the remaining nodes use $x_1 = 1$, $x_2 = 1.2$. 

There are three processor unit resource configurations. The configurations 1, 2, and 3 include 1, 2, and 4 instances of each type $CPU_0$, $CPU_1$, $GPU_0$, and $GPU_1$, respectively.

The $\mathit{MSWCRT}_{X}$ denotes the unsafe measured worst-case response time by online simulated execution of the scheduling algorithm X. And $\mathit{WCRT}_{X}$ refers to the safe WCRT derived by all-WCETs execution offline. The $\mathit{AVRT}_{X}$ refers to the average response time from online execution of X.

% It is important to note that not all generated systems, including constructors of DAGs, task execution times, and the number of processor units, will lead to TAs under a given scheduling algorithm. 

\textbf{Probability of TAs.} Before comparing response time metrics, we first analyze the probability of TAs for the systems under the HBFS and HFCFS with various configurations of DAGs and each configuration creates 100 DAGs all under the resource configuration 2 (a total of 10000 systems). We use a conservative criterion to determine if a system occurs TAs: if, after 10000 online simulation executions (with task execution times chosen randomly within their respective execution intervals)  under HBFS or HFCFS, any response time exceeds the one from the all-WCETs offline simulation, we consider the system occurs TAs. Tables~\ref{tab:HFCFS_TA_Pro} and~\ref{tab:HBFS_TA_Pro} show the probability of TAs, which indicate that as the interconnection probabilities increase, the likelihood of TAs decreases. This suggests that a reduced probability of inter-node parallelism lowers the likelihood of TAs. And, the probability of TAs occurring under HBFS is lower than HFCFS.

\begin{table}[t]
\centering
\caption{Probability(\%) of TAs under HFCFS Scheduling}
\vspace{-0.2cm}
\label{tab:HFCFS_TA_Pro}
\begin{tabular}{c cccccccc}
\toprule
 & \multicolumn{8}{c}{\textbf{Node Connectivity Probability}} \\
\cmidrule(lr){2-9}
\textbf{Node Number} & \textbf{0.1} & \textbf{0.2} & \textbf{0.3} & \textbf{0.4} & \textbf{0.5} & \textbf{0.6} & \multicolumn{2}{c}{\textbf{0.7-0.9}} \\
\midrule
\textbf{10} & 12.9 & 10.8 & 7.4 & 4.1 & 1.9 & 0.7 & \multicolumn{2}{c}{0.3} \\
\textbf{20} & 44.8 & 34.4 & 17.3 & 7.4 & 2.4 & 0.9 & \multicolumn{2}{c}{0.2} \\
\textbf{30} & 41.2 & 30.1 & 11.9 & 4.4 & 1.7 & 0.3 & \multicolumn{2}{c}{0.1} \\
\textbf{40} & 30.7 & 20.6 & 5.8 & 1.2 & 0.2 & 0.1 & \multicolumn{2}{c}{0} \\
% \textbf{50} & 6 & 6 & 3 & 0 & 0 & 0 & \multicolumn{2}{c}{0} \\
\bottomrule
\end{tabular}
\vspace{-0.3cm}
\end{table}

\begin{table}[t]
\centering
\caption{Probability(\%) of TAs under HBFS Scheduling}
\vspace{-0.2cm}
\label{tab:HBFS_TA_Pro}
\begin{tabular}{c cccccccc}
\toprule
 & \multicolumn{8}{c}{\textbf{Node Connectivity Probability}} \\
\cmidrule(lr){2-9}
\textbf{Node Number} & \textbf{0.1} & \textbf{0.2} & \textbf{0.3} & \textbf{0.4} & \textbf{0.5} & \textbf{0.6} & \multicolumn{2}{c}{\textbf{0.7-0.9}} \\
\midrule
\textbf{10} & 0.3 & 0.2 & 0.1 & 0.1 & 0.1 & 0 & \multicolumn{2}{c}{0} \\
\textbf{20} & 2.1 & 0.8 & 0.2 & 0.1 & 0 & 0 & \multicolumn{2}{c}{0} \\
\textbf{30} & 2.6 & 0.5 & 0.1 & 0 & 0 & 0 & \multicolumn{2}{c}{0} \\
\textbf{40} & 1.1 & 0.1& 0 & 0 & 0 & 0 & \multicolumn{2}{c}{0} \\
% \textbf{50} & 0 & 0 & 0 & 0 & 0 & 0 & \multicolumn{2}{c}{0} \\
\bottomrule
\end{tabular}
\vspace{-0.3cm}
\end{table}

\begin{table*}[t]
\caption{The average response time in systems with TAs}
% \vspace{-0.2cm}
\label{tab:withTA_only}
\centering
\begin{tabular}{c cccccccc}
\toprule
\textbf{$p$} 
& \scriptsize{\textbf{$\!\!\!\!\!\!\!\!ARA_{HFCFS}^{DDE_{HFCFS}}$}} & \scriptsize{\textbf{$\!\!\!\!\!\!\!\!SRA_{HFCFS}^{DDE_{HFCFS}}$}} & \scriptsize{\textbf{$\!\!\!\!\!\!\!\!ARA_{HBFS}^{DDE_{HBFS}}$}} & \scriptsize{\textbf{$\!\!\!\!\!\!\!\!SRA_{HBFS}^{DDE_{HBFS}}$}} 
& \scriptsize{\textbf{$\!\!\!\!\!\!\!\!ARA^{DDE_{HACAP}}_{HFCFS}$}} & \scriptsize{\textbf{$\!\!\!\!\!\!\!\!SRA^{DDE_{HACAP}}_{HFCFS}$}} & \scriptsize{\textbf{$\!\!\!\!\!\!\!\!ARA^{DDE_{HACAP}}_{HBFS}$}} & \scriptsize{\textbf{$\!\!\!\!\!\!\!\!SRA^{DDE_{HACAP}}_{HBFS}$}} \\
\midrule
\textbf{0.1} 
& 1.01 & 0.70 & 1.04 & 0.99 & 0.85 & 0.60 & 1.01 & 0.90 \\
\textbf{0.3} 
& 0.98 & 0.89 & 1.02 & 0.99 & 0.94 & 0.73 & 1.02 & 0.96 \\
\textbf{0.5} 
& 0.96 & 0.90 & 1.01 & 0.99 & 0.96 & 0.89 & 0.99 & 0.98 \\
\bottomrule
\end{tabular}
% \vspace{-0.3cm}
\end{table*}

\begin{table*}[t]
\centering
\caption{the average response time in random systems with or without TAs}
% \vspace{-0.2cm}
\label{tab:any_systems_only}

\begin{tabular}{c cccccccc}
\toprule
\textbf{$p$} 
& \scriptsize{\textbf{$\!\!\!\!\!\!\!\!ARA_{HFCFS}^{DDE_{HFCFS}}$}} & \scriptsize{\textbf{$\!\!\!\!\!\!\!\!SRA_{HFCFS}^{DDE_{HFCFS}}$}} & \scriptsize{\textbf{$\!\!\!\!\!\!\!\!ARA_{HBFS}^{DDE_{HBFS}}$}} & \scriptsize{\textbf{$\!\!\!\!\!\!\!\!SRA_{HBFS}^{DDE_{HBFS}}$}} 
& \scriptsize{\textbf{$\!\!\!\!\!\!\!\!ARA^{DDE_{HACAP}}_{HFCFS}$}} & \scriptsize{\textbf{$\!\!\!\!\!\!\!\!SRA^{DDE_{HACAP}}_{HFCFS}$}} & \scriptsize{\textbf{$\!\!\!\!\!\!\!\!ARA^{DDE_{HACAP}}_{HBFS}$}} & \scriptsize{\textbf{$\!\!\!\!\!\!\!\!SRA^{DDE_{HACAP}}_{HBFS}$}} \\
\midrule
\textbf{0.1} 
& 1.06 & 0.97 & 1.02 & 0.99 & 1.02 & 0.76 & 1.00 & 0.97 \\
\textbf{0.3}
& 1.02 & 0.97 & 1.01 & 0.99 & 1.04 & 0.82 & 1.08 & 0.95 \\
\textbf{0.5}
& 1.01 & 0.96 & 1.03 & 0.99 & 1.09 & 0.93 & 1.12 & 0.95 \\
\bottomrule
\end{tabular}
% \vspace{-0.3cm}
\end{table*}

% From Fig~\ref{muliteDAGcpm}, we observe that, after applying deterministic dynamic execution, the WCRT is consistently reduced. The line marked in blue and the line marked in red are both below 1. Furthermore, the $\mathit{WCRT}_{\text{HJACP}}$ produced by our proposed HJACP algorithm is almost always smaller than the other two, as shown in the bar chart, where most values are below 1. A bar height of 0 indicates that no timing anomalies were observed during testing. Notably, for DAGs with 20 nodes, the average $\mathit{WCRT}_{\text{HJACP}}$ is reduced by 15\% compared to $\mathit{WCRT}_{\text{HFCFS}}$. Additionally, applying deterministic execution to HFCFS yields greater WCRT reductions compared to HBFS. The method also performs better in cases with a small number of nodes and medium connectivity, or a large number of nodes and low connectivity.

%\textbf{Various Node Numbers and Interconnection Probabilities.}
\textbf{WCRT and MSWCRT.} 
We analyze the WCRT and MSWCRT for systems with TAs from the results of the experiment---probability of TAs. For systems where no TAs occur under algorithm X (HBFS or HFCFS), 
$WCRT_X$, $WCRT_{DDE_X}$, and $MSWCRT_X$ are same. As shown in Fig. 10, after applying DDE, the WCRT of X decreases compared to the MSWCRT of the original system, with values in the line plot consistently below 1. 
% On average, WCRT decreased by 5\%-25\%, with a maximum reduction of Y\%. 
The WCRT decreased by an average of 5\%-25\% across different DAG configurations. And in the best case, it was reduced by 72\%.
And DDE shows more improvements in systems with lower node interconnection probabilities, likely because TAs occur more frequently in such cases, providing greater optimization potential. Furthermore, when constraints generated by our HACPA algorithm are used, WCRT is further reduced in most cases, as indicated by the values in the bar chart being below 1. This demonstrates that constraints generated by HACPA help minimize WCRT. Missing bars or lines in the figure correspond to systems where no TAs systems were observed for that DAG structure, as detailed in Tables~\ref{tab:HFCFS_TA_Pro} and~\ref{tab:HBFS_TA_Pro}.

\begin{figure}
    \centering
    \includegraphics[width=1\linewidth]{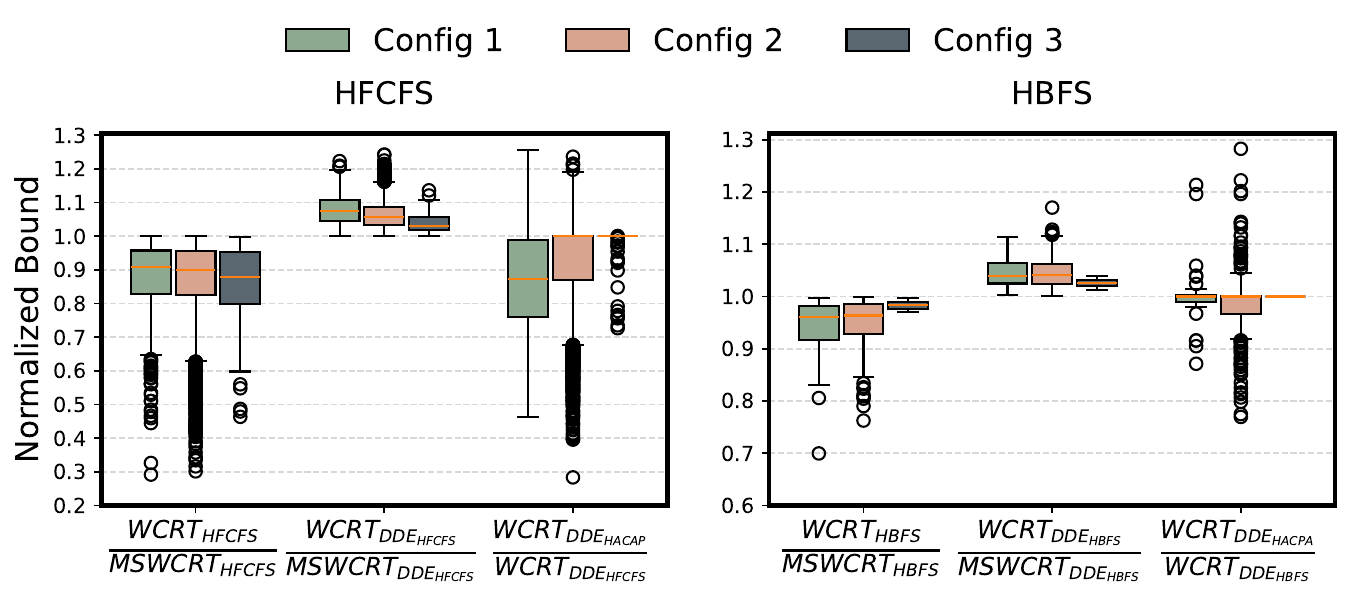}
    \caption{Comparison of different compute resource configurations}
    \label{fig:resource configuration}
\end{figure}

\textbf{Various Resource Configurations.} Here, we compare the performance of different scheduling algorithms under three resource configurations. Each DAG contains 10 to 30 nodes, and 100 DAGs are generated.  As a result, 10,000 systems with multi-type DAGs were generated for each resource configuration, still only analyzes the system with TAs.
As shown in Fig.~\ref{fig:resource configuration}, both HBFS and HFCFS exhibit TAs under multiple resource configurations (corresponding to the first group of box plots in each subfigure, all less than 1), whereas our DDF shows no TAs (corresponding to the second group of box plots, all bigger than 1). Moreover, the execution constraints generated by HACPA achieve overall better on reducing WCRTs on average (the third group of box plots). 
% It is worth noting that under resource configuration 3, HBFS exhibits almost no TAs, and thus no box plot.

\begin{figure}
    \centering
    \includegraphics[width=0.8
    \linewidth]{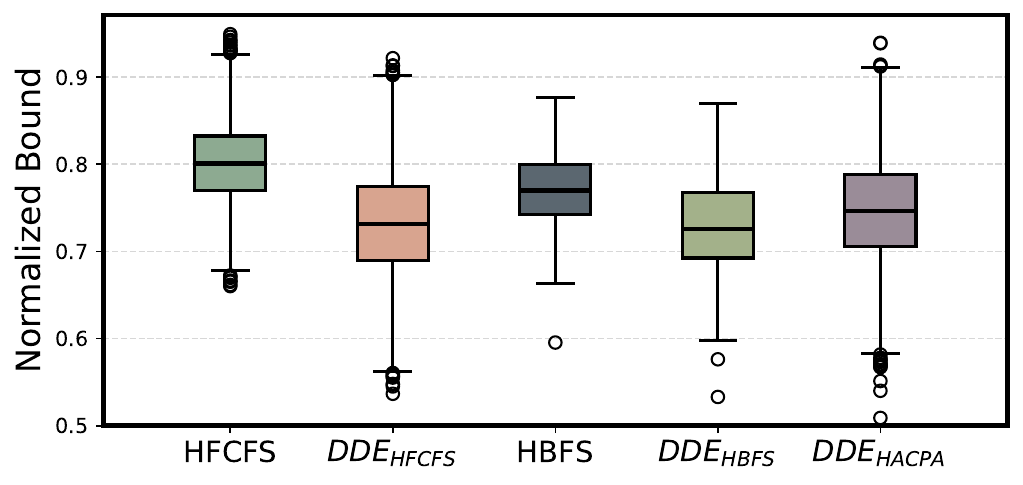}
    \caption{The response time jitter under different algorithms}   
    \label{fig:jitter_dif}
\end{figure}

\textbf{Response Time Jitter.} We analyze the response time jitter under resource configuration 2 with the result from the previous experiment.
For scheduling algorithm X, its jitter is defined as: $\frac{\mathit{MSWCRT}_{X} - \mathit{MSBCRT}_{X}}{\mathit{MSWCRT}_{X}}$.
As shown in Fig.~\ref{fig:jitter_dif}, the jitters of the DDE algorithms are significantly lower than those of traditional scheduling algorithms, with an average reduction of 7\%-9\%. This is because the DDE constrains part execution order and resource allocation, making the scheduling process more deterministic and effectively reducing jitters.

\textbf{Average Response Time.} We also compare the average response time (AVRT) of TAs systems (with TAs) and random systems (with or without TAs) in Table~\ref{tab:withTA_only} and ~\ref{tab:any_systems_only} among the different scheduling algorithms. The $RAVRT_{X,Y}$ is $ \frac{\mathit{AVRT}_{X}}{\mathit{AVRT}_{Y}}$. Furthermore, the $ARA_{X,Y}$ and $SRA_{X,Y}$ are the average and smallest of the ratios $RAVRT_{X,Y}$.
The results show that the DDE generally increases the average response time. Specifically, for typical systems (Table~\ref{tab:any_systems_only}), when the node interconnection probabilities are $\{0.1,0.3,0.5 \}$, the AVRT increase is 1\%-12\%, with an average increase of 4.2\%. However, not all systems experience an increase in AVRT; for columns of the $SRA$, the values are less than 1, indicating that the AVRT decreases.
For systems with TAs (Table~\ref{tab:withTA_only}), we even observe a reduction in AVRT in most cases. Under various interconnection probabilities, the AVRT decreases by 1.5\%, and in the best case, it decreases by up to 40\%. This improvement is likely because, in systems with TAs, our constraints not only optimize the worst-case execution but also eliminate certain scheduling spaces that could lead to poorer response times, thereby reducing the AVRT.

% $RA1$ and $RB1$ are based on the comparison between the deterministic execution of HJACP and the ordinary online execution of HFCFS. Similarly, $RAS2$, $RA2$, $RBS2$, and $RB2$ are the relevant cases for HBFS. The results show that the average performance compared to the original online algorithm after using deterministic constraints does not always decrease (greater than 1), especially in systems with TA, but it is in systems without TA that it generally decreases. Maybe it eliminates a range of worst-case scenarios. After using our HJACP algorithm, the average execution performance may improve in both systems with and without timing anomalies, especially for the HFCFS algorithm. It is possible that it identifies a constraint with higher parallelism.

\section{RELATED WORK}

Since Graham first pointed out in 1966 that task scheduling in multiprocessor systems could lead to timing anomalies (TAs)~\cite{graham1969bounds}, TAs have become a significant research issue.

In terms of system response time analysis, existing studies primarily focus on homogeneous systems. Some works~\cite{TAfreeWCRT1,TAfreeWCRT2,TAfreeWCRT3,TAfreeWCRT4} have proposed analytical methods to estimate the WCRT under conditions where TAs may occur. These approaches typically assume that all tasks on the critical path suffer from the worst possible interference from non-critical path tasks, resulting in overly conservative estimations. To avoid the occurrence of TAs, several studies have designed TA-free dynamic scheduling strategies, such as the LAZY scheduler~\cite{TAFree3}, the virtual-execution-based scheduler proposed by Chen~\cite{TAFree1}, and the DynFed scheduler~\cite{TAFree2}. And rencently, Lin et al.~\cite{TAFree4} explored the TAs in LF program scheduling

At the microarchitectural level, TAs in WCET analysis were first observed in dynamically scheduled processors by Lundqvist and Stenström in 1999~\cite{818824}. Subsequent research~\cite{9904748,binder2021still,reineke2006definition, cassez2012timing} delved into the definitions and causes of TA from the perspectives of micro-operations~\cite{reineke2006definition} and execution traces~\cite{cassez2012timing}, although a unified theoretical model that encompasses all types of anomalies is still lacking. In terms of analysis methods, some studies have proposed WCET analysis frameworks that tolerate TA~\cite{WCET2, WCET3}, while others have utilized structures such as the execution dependence diagram (XDD)~\cite{XDD1, XDD2, XDD3} to mitigate state space explosion caused by TAs. From the hardware perspective~\cite{cache2,MINOTAuR1,MINOTAuR2,MINOTAuR3,SIC1}, architectures with TA immunity, such as the SIC processor~\cite{SIC1}, have been designed to eliminate TAs at the hardware level.

Although the analysis of TAs at the microarchitectural level shares conceptual similarities with the challenges at the task scheduling level, there is currently no mature method that can be directly applied to heterogeneous systems. Therefore, constructing analyzable, predictable, and TA-free scheduling mechanisms for heterogeneous systems remains an urgent and open research challenge, which is the main work of this paper.
\section{CONCLUSION}
This paper presents the first work on implementing a timing-anomaly-free dynamic scheduling algorithm for heterogeneous systems—Deterministic Dynamic Execution, which enforces constraints on both task-to-resource mapping and execution order at runtime. The correctness of these constraints is formally proven. 
The experimental results show that our method effectively eliminates TAs while reducing both WCRT and response time jitter. For systems with or without TAs, the average response time shows only a slight increase. However, in systems with TAs, the average response time even decreases.

% \section*{Acknowledgments}
% This should be a simple paragraph before the References to thank those individuals and institutions who have supported your work on this article.

{\appendix[]

Proof of Lemma 1: Forward Progress Property.

 \begin{proof}
We first prove $c \sqsubseteq_C upd(c)$ holds, i.e., $\forall c \in C, \forall t \in \mathbb{T}, c(t) \sqsubseteq_P upd(c)(t)$. According to the definition of the state transition function $upd$ in Eqn.~\ref{Eq_upd:udp}, if task $t$ transitions to a new stage, it implies $c(t) \sqsubset_P upd(c)(t)$. Otherwise, the task remains in its current stage, and since the return value of  $tick'$ in Eqn.~\ref{Eq_tick_} is non-increasing, we still have $c(t) \sqsubseteq_P upd(c)(t)$. So the relation holds for all tasks, and we have $c \sqsubseteq_{C} upd(c)$.

Next, we prove that $c \sqsubset_{C} upd(c)$ holds. It suffices to prove that there exists at least one task $t' \in \mathbb{T}$ such that $c(t') \sqsubset_P upd(c)(t')$. Without loss of generality, let $t'$ be the first task in the execution order $O_D$ that has not yet entered the \texttt{Finish} stage. Since all tasks preceding $t'$ in $O_D$ have already completed, we have $ \texttt{Ready} \sqsubseteq_S c.\text{stage}(t') \sqsubset_S \texttt{Finish}$. Now consider the possible stages of $t'$:

\begin{itemize}
    \item If \( t' \) is in the \texttt{Ready} stage, then by the execution order constraint (constraint 2), all prior tasks have finished, and the succeeding tasks have not started. Thus, \( t' \) is guaranteed to be assigned a processing unit and enter \texttt{Exec} stage in the next transition, i.e., \( tran(t') = \text{true} \). It implies \( c.stage(t') \sqsubset_S upd(c).stage(t') \), hence \( c \sqsubset upd(c) \).  
    \item If \( t' \) is already in the \texttt{Exec} stage, then in the next state it will either move to \texttt{Finish}, or remain in \texttt{Exec} but with a reduced remaining time. In both cases, we have \( c(t_k) \sqsubset_P upd(c)(t_k) \), which again implies \( c \sqsubset upd(c) \).
\end{itemize}
Hence, Lemma 1 is proven.
\end{proof}

Proof of Lemma 2: Monotonicity of Dependency Removal.

\begin{proof}
If \( c_1.depComp(t) = \text{true} \), it means that all the tasks that task \( t \) depends on have entered the \texttt{Finish} stage in state \( c_1 \).  
Given \( c_1 \sqsubseteq c_2 \), we have  $\forall t \in T_d: \quad c_1.\text{stage}(t) \sqsubseteq_S c_2.\text{stage}(t) $ which implies that every task that has reached \texttt{Finish} in \( c_1 \) must also be in \texttt{Finish} in \( c_2 \). Therefore, in state \( c_2 \), all predecessors of task \( t \) have also completed, and hence \( c_2.depComp(t) = \text{true} \).
\end{proof}

Proof of Lemma 3: Monotonicity of Execution Stage.

To prove this condition, we can classify the cases based on the execution progress of task \( t \) in states \( c_1 \) and \( c_2 \). The proof can be divided into two main cases:

\begin{itemize}
    \item When \( c_1(t) = c_2(t) \), we further consider the four possible stages from \( S \) in Eqn.~\ref{Eq_stage}, and for each case, analyze the possible stage of task \( t \) in the next state.
    
    \item When \( c_1(t) \sqsubset c_2(t) \), we follow a similar case-based analysis, but additionally consider whether task \( t \) remains in the same stage in both \( c_1 \) and \( c_2 \).
\end{itemize}

Due to space limitations, we only provide a detailed proof for the representative case  
$c_1(t) = c_2(t) \land c_1.stage(t) = c_2.stage(t) = \texttt{Block} \land upd(c_1).stage(t) = \texttt{Exec},$  while the remaining cases can be handled similarly.  
We aim to prove that \( \texttt{Exec} \sqsubseteq_S upd(c_2).stage(t) \) holds.  
According to the definition of the transition function, e.g., Eqn.~\ref{Eq_tran}, this is equivalent to proving the following condition holds:  $c_2.tick(t) \leq 1 \land c_2.depComp(t) = true \land c_2.resAble(t) = true$.

\begin{proof}
Now, we prove that the three conditions hold.
% \( c_2.\text{tick}(t) \leq 1 \), \( c_2.\text{depComp}(t) = \text{true} \) and \( c_2.\text{resAble}(t) = \text{true} \).
\begin{itemize}
    \item \textbf{Prove that \( c_2.tick(t) \leq 1 \):}  
    Since task \( t \) transitions to a new stage in \( upd(c_1) \), it must satisfy \( c_1.tick(t) \leq 1 \).  
    Given that \( c_1(t) = c_2(t) \), it follows directly that \( c_2.tick(t) \leq 1 \).

    \item \textbf{Prove that \( c_2.depComp(t) = \text{true} \):}  
    This follows directly from Lemma 2.

    \item \textbf{Prove that \( c_2.resAble(t) = \text{true} \):}  We proceed by contradiction, which assumes that task \( t \) in the state $upd(c_2)$, does not have any available processor unit satisfying the resource constraint from a given $resAllocPat_D$.  

    Let \( res_t \) denote the type for  \( t \) from $resAllocPat_D$.  
    Suppose in the state  \( udp(c_2) \), there exists a set \( \mathbb{K}_1 \) with \( k_1 \) tasks that will still occupy resources of type \( res_t \) (i.e., the set described by $c_2.nextStageStillUse(res_t)$),  
    and a set \( \mathbb{K}_2 \) with \( k_2 \) tasks that will request resources of type \( res_t \), with higher priority than \( t \) in a given order $O_D$ from constraint 2, and are currently in the \texttt{Ready} stage (i.e., described by Eqn.~\ref{Eq_hiTaskNum4Res}).  
    So, we have: $k_1 + k_2 \geq resNum(res_t)$, due to the assumption.

    Since \( c_1 \sqsubseteq c_2 \), the progress of all tasks in \( \mathbb{K}_1 \cup \mathbb{K}_2 \) in \( c_1 \) is no greater than in \( c_2 \). Thus:$(\forall t' \in \mathbb{K}_1,\; c_1(t') \sqsubseteq_P c_2(t') \sqsubset_P (\texttt{Exec}, 1)) \quad \land \quad (\forall t' \in \mathbb{K}_2,\; c_1.stage(t') \sqsubseteq_S c_2.stage(t') = \texttt{Ready}) $.

    In state \( c_1 \), all tasks in \( \mathbb{K}_1 \) are at most in the \texttt{Exec} stage. Since task \( t \) successfully transitions to \texttt{Exec} in \( upd(c_1) \), the \( k_1 \) tasks must be in \texttt{Exec} in the next state as well.  
    Moreover, since task \( t \) is granted a procosser unit in \( upd(c_1) \), and tasks in \( \mathbb{K}_2 \) have higher priority, they must also be able to obtain procosser units. 
    Due to Constraint 1, in all states the task from $\mathbb{K}_1$  and $\mathbb{K}_2$ use procosser unit type $res_t$, which implies  \( k_1 + k_2 + 1 \leq resNum(res_t) \), contradicting our earlier assumption.  
    Therefore, our assumption is false, and \( c_2.resAble(t) = \text{true} \) must hold.
\end{itemize}
Hence, Lemma 3 is proven.
\end{proof}

 }


\bibliographystyle{IEEEtran}
\bibliography{reference}

\newpage
\section{Biography Section}

% \bf{If you include a photo:}
\vspace{-20pt}
\begin{IEEEbiography}[{\raisebox{0.25\height}{\includegraphics[width=1in,height=1.25in,clip,keepaspectratio]{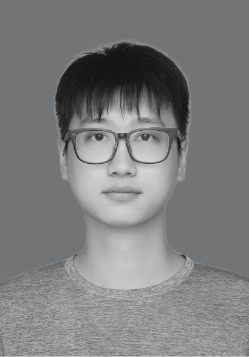}}}]{Yixuan Zhu}
received the B.S. degree in software engineering from the University of Electronic Science and Technology of China, Chengdu, China, in 2022. He is currently pursuing a Ph.D degree in computer science at the University of Science and Technology of China, Hefei, China. His research interests include time behavior analysis, microarchitecture modeling, and timing anomalies.
\end{IEEEbiography}

\vspace{-20pt}

\begin{IEEEbiography}[{\raisebox{0.25\height}{\includegraphics[width=1in,height=1.25in,clip,keepaspectratio]{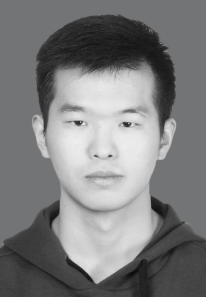}}}]{Yinkang Gao}received the B.S. degree in computer science from the University of Nanjing University of Posts and Telecommunications, Nanjing, China, in 2020. He is currently pursuing a Ph.D degree in computer science at the University of Science and Technology
 of China, Hefei, China. His research interests include computer architectures and embedded systems.
\end{IEEEbiography}

\vspace{-20pt}

\begin{IEEEbiography}[{\raisebox{0.25\height}{\includegraphics[width=1in,height=1.25in,clip,keepaspectratio]{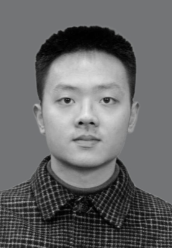}}}]{Binze Jiang}
received his master's degree in computer science and technology from the University of Science and Technology of China, China, in 2025. His research interests include computer architecture and real-time systems.
\end{IEEEbiography}

\vspace{-20pt}

\begin{IEEEbiography}[{\raisebox{0.25\height}{\includegraphics[width=1in,height=1.25in,clip,keepaspectratio]{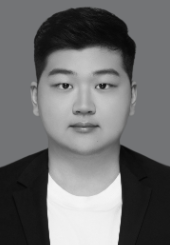}}}]{Xiaohang Gong}
received the B.E. degree in computer science and technology in 2022 and the M.E. degree in computer science and technology in 2025 from the University of Science and Technology of China (USTC), Hefei, China.  His research interests include real-time systems, time-predictable cache coherent systems, and memory system design.
\end{IEEEbiography}

\vspace{-20pt}

\begin{IEEEbiography}[{\raisebox{0.25\height}{\includegraphics[width=1in,height=1.25in,clip,keepaspectratio]{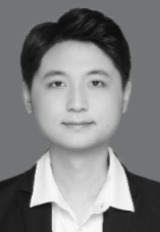}}}]{Zihan Wang}
received the B.S. degree in computer science and technology from Shandong University, Qingdao, China, in 2022. He is currently working toward the Ph.D. degree in computer science at the University of Science and Technology of China, Hefei, China. His research interests include AI compilers, sparse computation, and machine learning systems.
\end{IEEEbiography}

\vspace{-20pt}

\begin{IEEEbiography}[{\raisebox{0.25\height}{\includegraphics[width=1in,height=1.25in,clip,keepaspectratio]{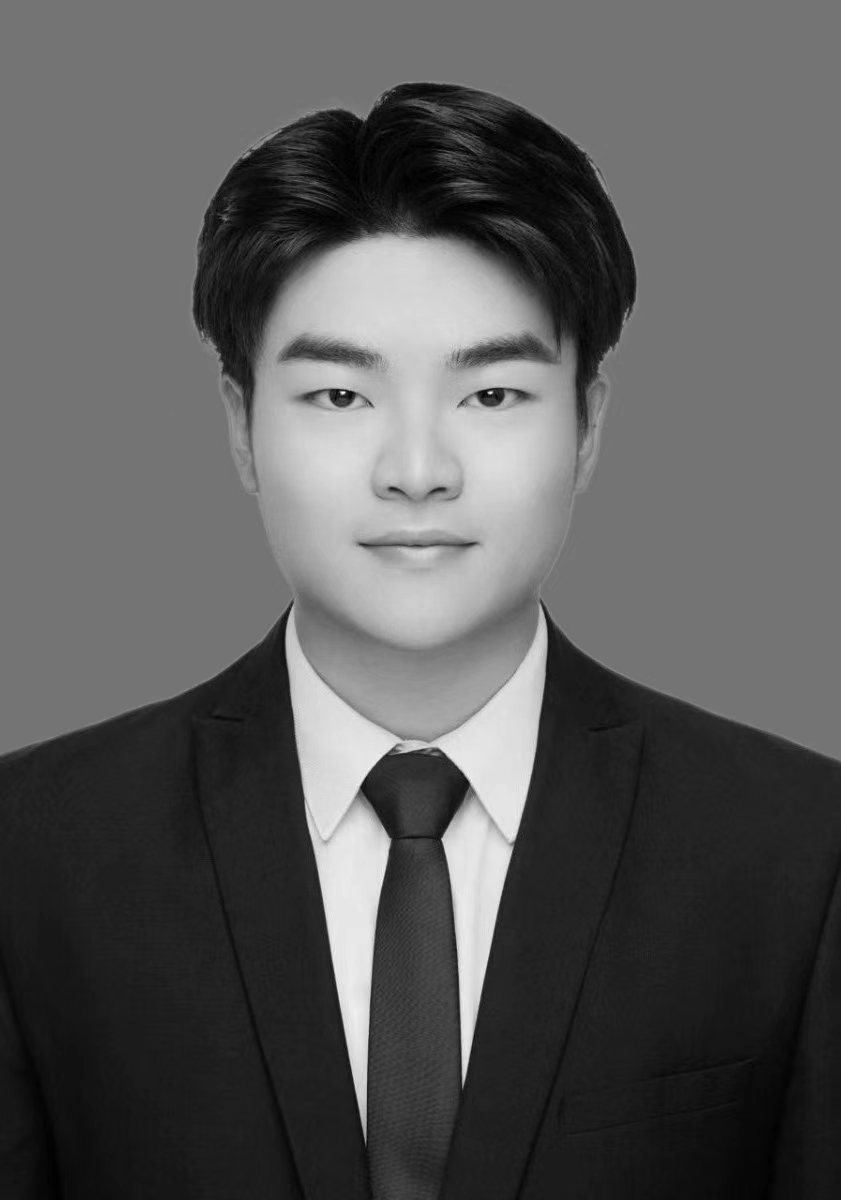}}}]{Cheng Tang}
is a Ph.D. candidate in Software Engineering at the University of Science and Technology of China. His research interests lie in efficient AI algorithms, especially inference acceleration for large language models, including memory-efficient inference, runtime optimization, and hardware-aware system design.
\end{IEEEbiography}

\vspace{-20pt}

\begin{IEEEbiography}[{\raisebox{0.0\height}{\includegraphics[width=1in,height=1.25in,clip,keepaspectratio]{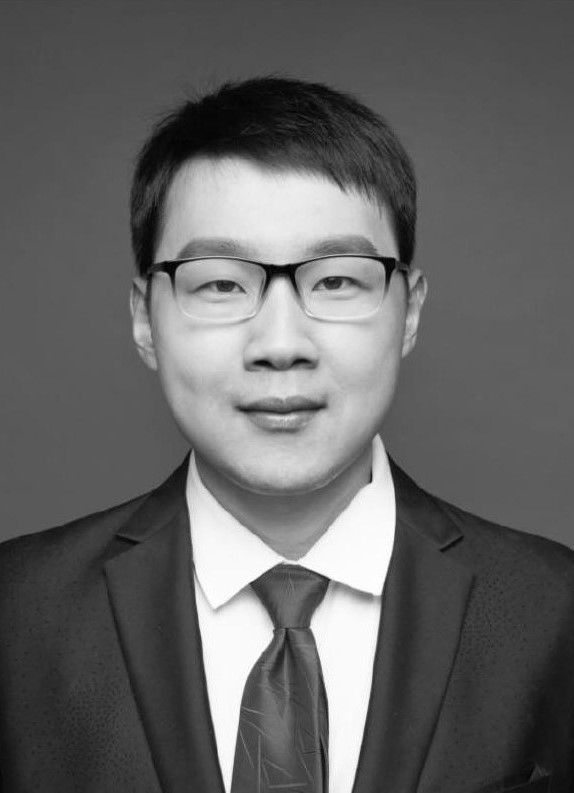}}}]{Lei Gong}
received the Ph.D. degree in computer science from the University of Science and Technology of China, Hefei, China, in 2019. He is an Associate Professor with the School of Computer Science, University of Science and Technology of China. His research interests include FPGA-based accelerator designs and artificial intelligence and machine learning systems. Dr. Gong’s paper was once nominated as Best Paper Candidate in CODES+ISSS 2018.
\end{IEEEbiography}

\vspace{-20pt}

\begin{IEEEbiography}[{\raisebox{0.0\height}{\includegraphics[width=1in,height=1.25in,clip,keepaspectratio]{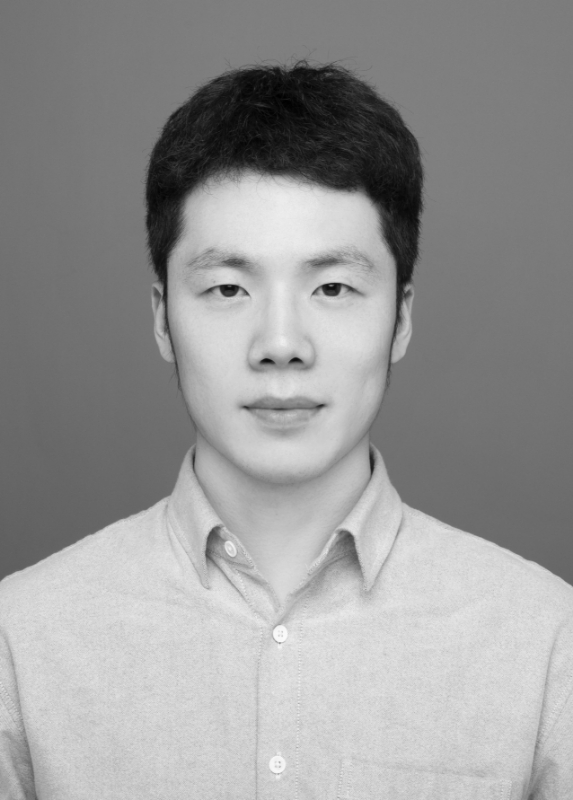}}}]{Wenqi Lou}
received the B.S. degree from Northwestern Polytechnical University, Xi’an, China, in 2018, and the Ph.D. degree in computer science from the University of Science and Technology of China, Hefei, China, in 2023. He is currently an Associate Researcher with the School of Software Engineering, University of Science and Technology of China. His current research interests include deep learning accelerators and FPGA-based acceleration.
\end{IEEEbiography}

\vspace{-5pt}

\begin{IEEEbiography}[{\raisebox{0.2\height}{\includegraphics[width=1in,height=1.25in,clip,keepaspectratio]{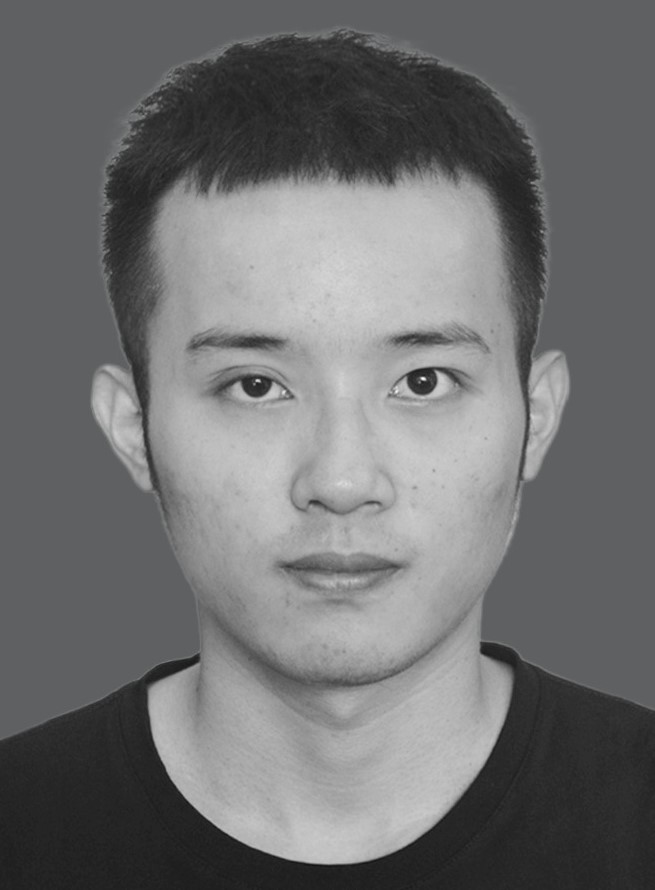}}}]{Teng Wang}
received the Ph.D. degree from the University of Science and Technology of China, Hefei, China, in 2023. He is currently a Research Scientist with Suzhou Institute for Advanced Research, University of Science and Technology of China, Suzhou, China. His research interests include algorithm-level and architecture-level optimizations of FPGAs for deep learning applications.
\end{IEEEbiography}

\vspace{-15pt}

\begin{IEEEbiography}[{\raisebox{0.0\height}{\includegraphics[width=1in,height=1.25in,clip,keepaspectratio]{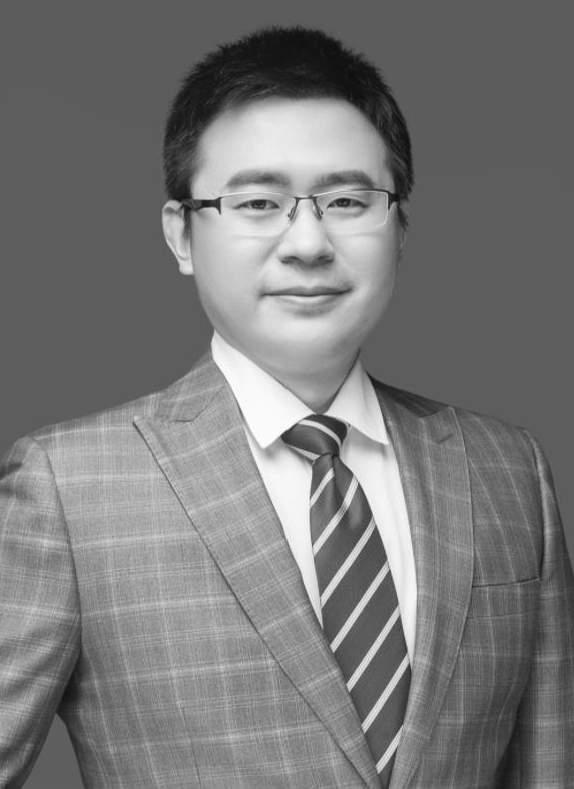}}}]{Chao Wang}
(Senior Member, IEEE) received the B.S. and Ph.D. degrees in computer science from the University of Science and Technology of China, Hefei, China, in 2006 and 2011, respectively. He is currently a Professor with the University of Science and Technology of China. His research interests include multicore and reconfigurable computing. Prof. Wang serves as the Associate Editor for ACM Transactions on Design Automation for Electronics Systems and IEEE/ACM Transactions on Computational Biology and Bioinformatics.
\end{IEEEbiography}

\vspace{-15pt}

\begin{IEEEbiography}[{\raisebox{0.0\height}{\includegraphics[width=1in,height=1.25in,clip,keepaspectratio]{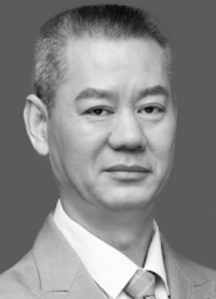}}}]{Xi Li}
received the PhD degree in computer science from the University of Science and Technology of China in 2003, and is currently a professor in the School of Computer Science and Technology, and the School the Software Engineering, University of Science and Technology of China. There, he directs the research programs in the High Energy-efficiency Intelligent Computing Lab, examining various aspects of computer systems, especially real-time embedded systems.
\end{IEEEbiography}

\vspace{-15pt}

\begin{IEEEbiography}[{\raisebox{0.0\height}{\includegraphics[width=1in,height=1.25in,clip,keepaspectratio]{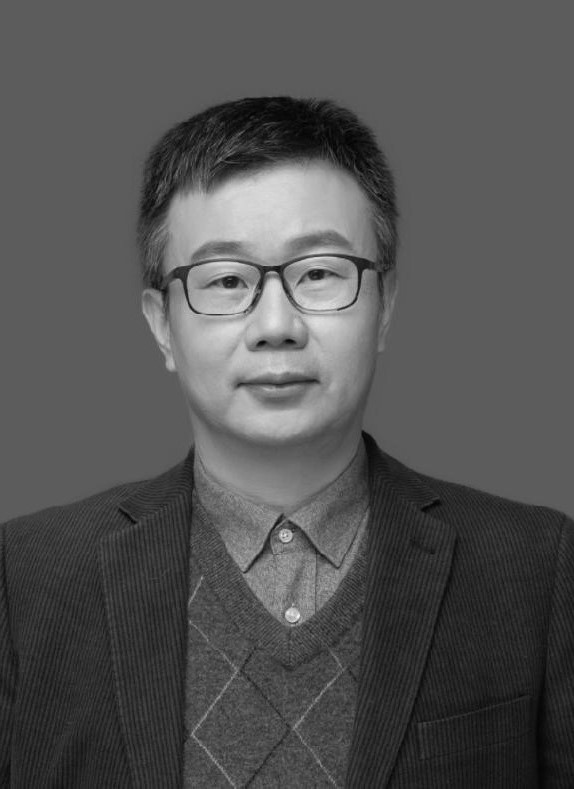}}}]{Xuehai Zhou}
received the B.S., M.S., and Ph.D. degrees from the University of Science and Technology of China, Hefei, China, in 1987, 1990, and 1997, respectively. He is currently a Professor with the School of Computer Science, University of Science and Technology of China. He serves as a General Secretary of the Steering Committee of Computer College Fundamental Lessons and the Technical Committee of Open Systems, China Computer Federation.
\end{IEEEbiography}

\vfill

\end{document}